%% file: main.tex
\documentclass[3p, preprint]{elsarticle}

\usepackage[hidelinks]{hyperref}

\journal{Journal of Logical and Algebraic Methods in Programming}

\usepackage[T1]{fontenc}
\usepackage[utf8]{inputenc}
\usepackage{xcolor}
\usepackage{multirow}
\biboptions{comma,numbers,sort}

\usepackage{tikz} % tikz figures
\usetikzlibrary{trees} % tikz trees
\usetikzlibrary{automata,positioning} % tikz automata
\usepackage{pgfplotstable} % plots
\usepackage{pgfplots} % plots

\usepackage[activate={true,nocompatibility},final,tracking=true,kerning=true,spacing=true,factor=1100,stretch=10,shrink=10]{microtype}

\usepackage{
  subcaption,
  lmodern,
  listings,
  amsmath,
  paralist
}

\lstdefinestyle{capabilities}{
  language=java,
  basicstyle=\fontsize{11}{12}\selectfont\tt\color{black},
  keywordstyle=\fontsize{11}{12}\selectfont\bf,
  numberstyle=\fontsize{5}{10}\selectfont\tt\color{black},
  commentstyle=\color{magenta}\it,
  aboveskip=1ex,
  belowskip=1ex,
  tabsize=2,
  columns=fullflexible,
  xleftmargin=1ex,
  resetmargins=true,
  showstringspaces=false,
  morecomment=[l]{//},
  morecomment=[l]{--},
  morecomment=[s]{/*}{*/},
  escapeinside=??,
  morekeywords={actor,active,acquire,and,atomic,release,or,not,assert,bestow,unless,B,A,def,var,val,subordinate,let,in,end,,then,else,linear,subord,Unit, match, with, safe, await},
  moredelim=[is][\textit]{___}{___},
  moredelim=[is][\textbf]{__*}{*__}
}

\lstnewenvironment{code}{\lstset{style=capabilities}}{}
\lstnewenvironment{codep}[1][numbers=left]{\lstset{style=capabilities,#1}}{}
\lstnewenvironment{codex}[1][numbers=left]{\lstset{style=capabilities,#1}}{}

\newcommand{\ie}{\emph{i.e.,}}
\newcommand{\eg}{\emph{e.g.,}}
\newcommand{\cf}{\emph{cf.,}}
\newcommand{\etal}{\emph{et~al.}}
\newcommand{\Pad}{\vspace*{1ex}}
\newcommand{\Tighten}{\vspace*{-1ex}}

\renewcommand{\c}[1]{\lstinline[style=capabilities,basicstyle=\fontsize{10}{10}\selectfont\tt,keywordstyle=\fontsize{10}{10}\selectfont\bf]@#1@}

\renewcommand{\bfdefault}{b}

\DeclareFontFamily{T1}{lmtt}{}
\DeclareFontShape{T1}{lmtt}{m}{n}{<-> ec-lmtl10}{}
\DeclareFontShape{T1}{lmtt}{m}{\itdefault}{<-> ec-lmtlo10}{}
\DeclareFontShape{T1}{lmtt}{\bfdefault}{n}{<-> ec-lmtk10}{}
\DeclareFontShape{T1}{lmtt}{\bfdefault}{\itdefault}{<-> ec-lmtko10}{}

\newcommand{\CF}[1]{\emph{cf.}~\SecRef{#1}}
\newcommand{\SecRef}[1]{Section~\ref{sec:#1}}
\newcommand{\FigRef}[1]{Figure~\ref{fig:#1}}

\newcommand{\SecLabel}[1]{\label{sec:#1}}
\newcommand{\FigLabel}[1]{\label{fig:#1}}

\usepackage{supertabular}
\usepackage[supertabular,lineBreakHack]{ottalt}

  \input{semantics.ott.tex}

  \renewottcommands[ott]

\newcommand{\RN}[1]{\textnormal{\textls{\uppercase{\scriptsize(#1)}}}}
\newcommand{\ARN}[1]{\textnormal{\textls{\uppercase{\scriptsize#1}}}}

\begin{document}
\begin{frontmatter}
%% == Title ======================================================
\title{Bestow and Atomic: Concurrent Programming using Isolation, Delegation and Grouping}

\tnotetext[funding]{This work is sponsored by the UPMARC center
  of excellence, the FP7 project ``UPSCALE'' and the project
  ``Structured Aliasing'' financed by the Swedish Research
  Council.}

\author{Elias Castegren\corref{mycorrespondingauthor}}
\ead{elias.castegren@it.uu.se}
\author{Joel Wallin}
\ead{joel.wallin.3149@student.uu.se}
\author{Tobias Wrigstad}
\ead{tobias.wrigstad@it.uu.se}
\address{Uppsala University, Sweden}
\cortext[mycorrespondingauthor]{Corresponding author}

\begin{abstract}
  Any non-trivial concurrent system warrants synchronisation, regardless of the concurrency model.
  Actor-based concurrency serialises \emph{all} computations in an
  actor through asynchronous message passing.
  In contrast, lock-based concurrency serialises \emph{some}
  computations by following a lock--unlock protocol for accessing
  certain data.

  Both systems require sound reasoning about pointers and aliasing
  to exclude data-races. If actor isolation is broken, so is
  the single-thread-of-control abstraction. Similarly for locks,
  if a datum is accessible outside of the scope of the lock, the
  datum is not governed by the lock.

  In this paper we discuss how to balance aliasing and
  synchronisation. In previous work, we defined a type system that
  guarantees data-race freedom of actor-based concurrency and
  lock-based concurrency. This paper extends this work by the
  introduction of two programming constructs; one for decoupling
  isolation and synchronisation and one for constructing
  higher-level atomicity guarantees from lower-level
  synchronisation. We focus predominantly on actors, and in
  particular the Encore programming language, but our ultimate
  goal is to define our constructs in such a way that they can be
  used both with locks and actors, given that combinations of both
  models occur frequently in actual systems.

  We discuss the design space, provide several formalisations of
  different semantics and discuss their properties, and connect
  them to case studies showing how our proposed constructs can be
  useful. We also report on an on-going implementation of our
  proposed constructs in Encore.
\end{abstract}

\end{frontmatter}

\section{Introduction}

Concurrency can be defined as coordinating access to shared
resources. Synchronisation is naturally a key aspect of concurrent
programs and different concurrency models handle synchronisation
differently. Pessimistic models, like locks or the actor model
\cite{hewitt1973session, baker1977laws} serialise computation \emph{within certain
  encapsulated units}, allowing sequential reasoning about
internal behaviour at the cost of sometimes pruning possible
parallel performance gains. In contrast, optimistic models, like
lock-free programming \cite{ArtOfMultiProg} or software transactional memory
\cite{Art:STM} allow concurrent operations on the same data, but
require that all operations follow some protocol to exclude
unwanted behaviour, or avoid side-effects that cannot be
rolled-back in the event of conflicts between threads. In this
paper, we focus on pessimistic models.

% Similarly, transactions must record every load
% and store inside a particular scope to be able to roll-back in the
% event of a conflict.

In the case of the actor model, if a reference to an actor $A$'s
internal state is accessible outside of $A$, operations inside of
$A$ are subject to data-races and sequential reasoning is lost.
The same holds true for operations on an aggregate object behind a
lock, if a sub-object is leaked and becomes accessible where the
appropriate lock is not held.

In previous work, we designed Kappa \cite{castegren16}, a type
system in which the boundary of a unit of encapsulation can be
statically identified. An entire encapsulated unit can be wrapped
inside some synchronisation mechanism, \eg{} a lock or an
asynchronous actor interface, and consequently all operations
inside the boundary are guaranteed to be data-race free. An
important goal of this work is facilitating object-oriented reuse
in concurrent programming: internal objects are oblivious to how
their data-race freedom is guaranteed, and the building blocks can
be reused without change regardless of their external
synchronisation. Further, making synchronisation tractable
simplifies concurrent programming as the
portions of a system that are accessed concurrently will be identified,
and compilers can verify that the program behaves in accordance with
the programmer's intention, with respect to concurrent accesses.

This paper explores two extensions to the Kappa system, which we
explain in the context of the actor model (although they are
equally applicable to a system using locks). The first extension,
\emph{bestow}, allows references to an object to escape its
\emph{unit of encapsulation} without escaping its \emph{unit of
  synchronisation}; all external operations on private state will
be implicitly delegated to the owner of that state, either via
message passing, or by acquiring a lock specific to the owner of
the private state.
This enables several useful programming
patterns without relaxing Kappa's static data-race freedom
guarantee. For example, in the context of an actor system, actors
may safely leak references to state, effectively allowing many
objects to cooperate in constructing the actor's interface.
Similarly, in the context of a lock, it will be possible to hold
on to references deep inside a structure, even when the lock is
not held, with a guarantee from the type system that these will not be used until the
lock is re-acquired.

When encapsulation and synchronisation are decoupled, another
extension becomes necessary to group operations together,
enabling \emph{atomicity} of multiple operations on a data structure that potentially has several
different entry-points. To this end, we
introduce an \emph{atomic} block scoped construct. In the context
of an actor system, this allows \eg{} grouping several messages to
prune unwanted interleavings. In the context of a lock-based
system, it allows performing several distinct operations without releasing the lock in-between.

This paper makes the following contributions:

\begin{enumerate}
\item We discuss (\SecRef{DRF}) the difference between actor
  systems where data-races are avoided through \emph{isolation} of
  an actor's passive objects and actor systems where data-races are
  avoided through \emph{delegation} of operations on passive
  objects to the actors that own them, and show how both are
  supported in the Encore programming language \cite{encoreSFM}.
  Encore supports delegation through \emph{bestowed references},
  which were introduced in Encore as part of this work. In
  contrast to systems like E \cite{RobustComposition} and
  AmbientTalk \cite{AmbientTalk}, Encore's delegation is
  purposely \emph{not} transparent, allowing programmers to reason
  about the performance and latency of operations and distinguish
  between operations on local and remote objects (\SecRef{bestow}).

\item We extend the delegation concept with a notion of movement
  of objects between actors, \ie{} implicit and/or automated
  transfer of ownership. This enables a form of load-balancing of
  passive objects between actors (\SecRef{load-balancing},
  formalised in \SecRef{formal2}).

\item We introduce a block-scoped construct for grouping
  operations to be performed as an atomic unit that can be applied
  to both actor-based systems and lock-based systems, as well as
  systems that combine the two models (\SecRef{atomic}).

\item We explore and formalise three variations of the semantics of bestowed
  references and atomic blocks, and show that the resulting systems are free from data-races. For simplicity, our formalisations
  are based on simple $\lambda$-calculus models with actors. We provide
  mechanised versions in Coq for others to build on. (Sections
  \ref{sec:formal1}--\ref{sec:formal2}).

\item We report on a number of small case studies using our
  proposed constructs (\SecRef{case-studies}) and on the on-going
  implementation of bestowed references and atomic blocks in
  Encore (\SecRef{implementation}).

\end{enumerate}

This paper extends earlier work~\cite{Art:Bestow} by adding the
two variations of the semantics (\SecRef{formal2}), mechanising
the semantics and their proofs~\cite{bestow-repo}, providing case
studies (\SecRef{case-studies}), reporting on the current state of
implementation (\SecRef{implementation}), and adding more in-depth
discussions about the work throughout all sections of the paper.

\section{Background: Kappa and Encore}
\SecLabel{background}

This section covers the background needed to understand the
context of this work. It explains the basics of Kappa, a type
system which guarantees data-race freedom in concurrent programs,
and Encore, an actor-based language which uses Kappa to facilitate
safe sharing between actors.

% \subsection{Kappa in a Nutshell}

Kappa is a type system for concurrent object-oriented
programming~\cite{castegren16}. Its most important guarantee is
that no two concurrent operations will access the same memory
address, unless both operations only use this address for reading.
Kappa achieves this by preventing the creation of aliases which
could be used to cause data-races. A reference can only be shared
between threads if all accesses are synchronised (\eg{} by using
locks) or if all operations available through the reference are
non-mutating.

The sharing properties of a reference are specified by its
\emph{mode}, which is tracked by the type system. For example, a
reference with the \emph{locked} mode is implicitly protected by a
lock, similar to a Java object whose methods are all
\c{synchronized}. It may be safely shared between threads, as any
concurrent accesses will be synchronised. In contrast, a reference
with the \emph{subordinate} mode may not be shared between
threads. In fact, it may not even leak outside of the object that
created it (similar to ownership types \cite{OT}). This means that
a locked object (whose aliases all have the locked mode) may
encapsulate additional subordinate objects. Since these objects
may only be accessed via the locked object (thus grabbing its
lock), they can be operated on without additional synchronisation.

% \subsection{Encore in a Nutshell}

Kappa has been implemented as the type system for Encore, an
object-oriented language developed in the context of the UPSCALE
project~\cite{encoreSFM}. Encore uses actors to achieve
concurrency, and Kappa ensures that objects shared between actors
are always accessed without data-races. Encore's actors are
implemented by extending Kappa with an \emph{actor} mode.
References with this mode point to actors and may only be used to
send messages. Since all interaction with an actor go via its
message queue, operations on it will be synchronised, similar to
the operations on a locked object. Just like locked objects,
actors can use subordinate references to ensure that their own
private state is never accessed by other actors. Additionally,
Encore allows sharing of immutable or read-only data, and supports
ownership transfer of objects.

The interface of an actor is defined by its class, and the
messages in an actor's message queue are processed in sequence
(there is no selective receieving of messages, as in \eg{}
Erlang~\cite{erlang}). This means that Encore's actors behave as
active objects~\cite{activeObjects}. The work relating to actors
in this paper was done with Encore's actors in mind, but is
applicable to any actor language with active object semantics.

The constructs presented in this paper rely both on the
encapsulation guarantees given by the subordinate mode, as well as
the synchronisation guarantees given by locks and actors. They are
however orthogonal to which synchronisation technique is used (and
can be used in a system that uses both).
Encore does not yet support Kappa's locked mode because of
the complexity of integrating locks with Orca
\cite{Art:ORCAOOPSLA}---Encore's garbage collection protocol (which
is shared with the Pony programming language \cite{ponyAgere}).
Besides implementation, garbage collection is orthogonal to the
matters discussed in this paper.

\section{Data-Race Freedom: Delegation and Isolation}
\SecLabel{DRF}

Actors simplify concurrent programming by enabling sequential
reasoning inside each actor. The majority of actor-based
programming languages and frameworks---\eg{} Akka \cite{Akka},
ProActive \cite{ProActive}, ABS \cite{ABS}, Orleans
\cite{bernstein2014orleans}, Encore \cite{encoreSFM}, Pony
\cite{ponyAgere} and Joelle \cite{Joelle,theoestlund} rely on
\emph{isolation}, enforced manually by programmer diligence or
through compiler support to do so: by preventing references to an
actor's state to leak outside of the actor, the only way to
manipulate an actor's state is by sending it a message. (Compilers
and run-time systems may either reject references which would
violate encapsulation, or insert instructions to deep-copy message
payloads.) \FigRef{illustrate} (left) shows the key situation to
be prevented to avoid data-races. The brown circles denote actors,
each with a (logical) thread of control. If the external reference
to \c{foo} is possible, accesses to foo could be subject to data
races.

An alternative model to isolation is found in some actor systems,
such as E \cite{RobustComposition} and AmbientTalk
\cite{AmbientTalk} which allows an actor's state to be
arbitrarily referenced, but requires that operations on the state of some other
actor is implicitly delegated to that actor. In the isolation
model, data-race freedom is achieved through enforcing that all
external access go via an asynchronous interface. In the
delegation model, an actor can be thought of having just one
method in its interface: \emph{perform}, and each operation is
translated into a perform call with some appropriate lambda.
Internal calls to perform can be carried out synchronously.
\FigRef{illustrate} (right) shows how operations on a passive
object belonging to another actor can be enabled by making the
operations asynchronous and asking the actor that owns the passive
object to carry them out. Thus, only the actor owning the object
$o$ pointed to by \c{foo} will ever operate on $o$.

The isolation model ties into existing theory on abstraction,
modularisation, and encapsulation. In most languages, it serves to
simplify reasoning: operations on passive objects are synchronous
and efficient whereas operations on actors are asynchronous and
introduce latency, but also make concurrency and parallelism
possible. The isolation model also helps clearly delimiting actor
interfaces. In contrast, the delegation model is much less
restrictive on the object graph of a system, but (in E and
AmbientTalk) complicates reasoning about the performance and
latency, and makes an actor's interface less clearly
delimited and defined.

Restricting object graph topologies means limiting aliasing in a
system. This is unsurprising since aliasing is a prerequisite of
data-races---but far from all aliasing is bad. Aliasing is
commonly used to provide shortcuts through a data structure, such
as a last pointer in a linked list implementation to allow
constant-time append operations. Staying with the linked list
example, we now demonstrate how delegation can be useful in an
actor setting. While the example is a little contrived, it
demonstrates the importance of allowing the creation of---and
sharing of---aliases.

\begin{figure}[t]
  \centering
  \includegraphics[width=.4\textwidth]{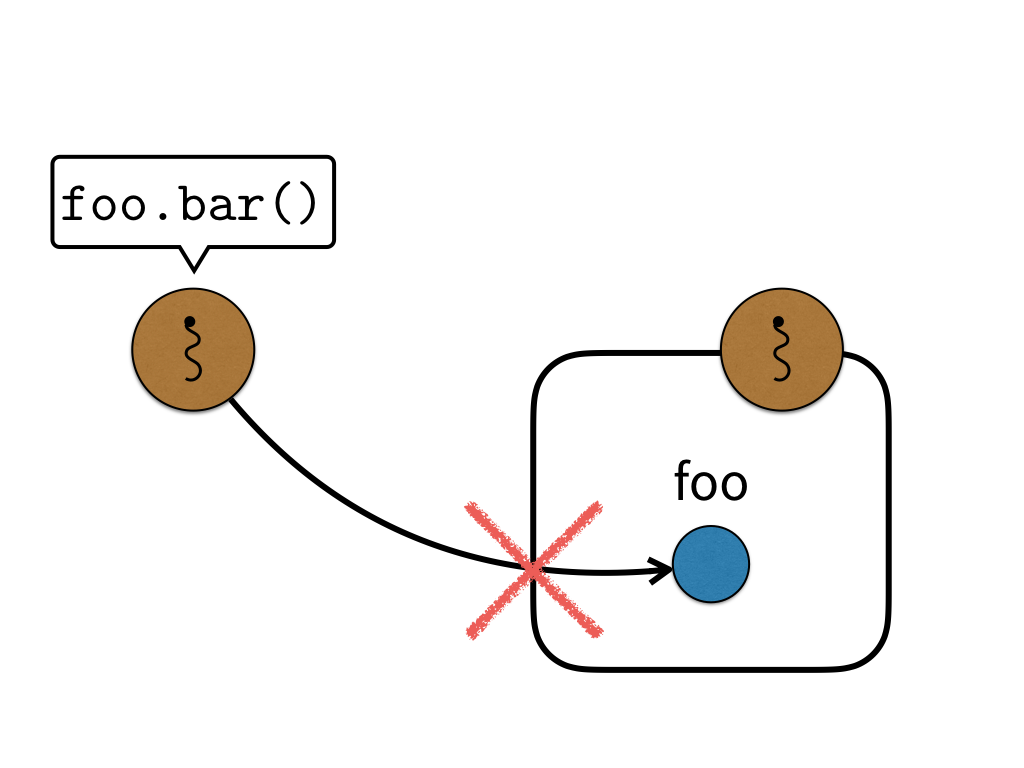}
  \qquad
  \qquad
  \includegraphics[width=.4\textwidth]{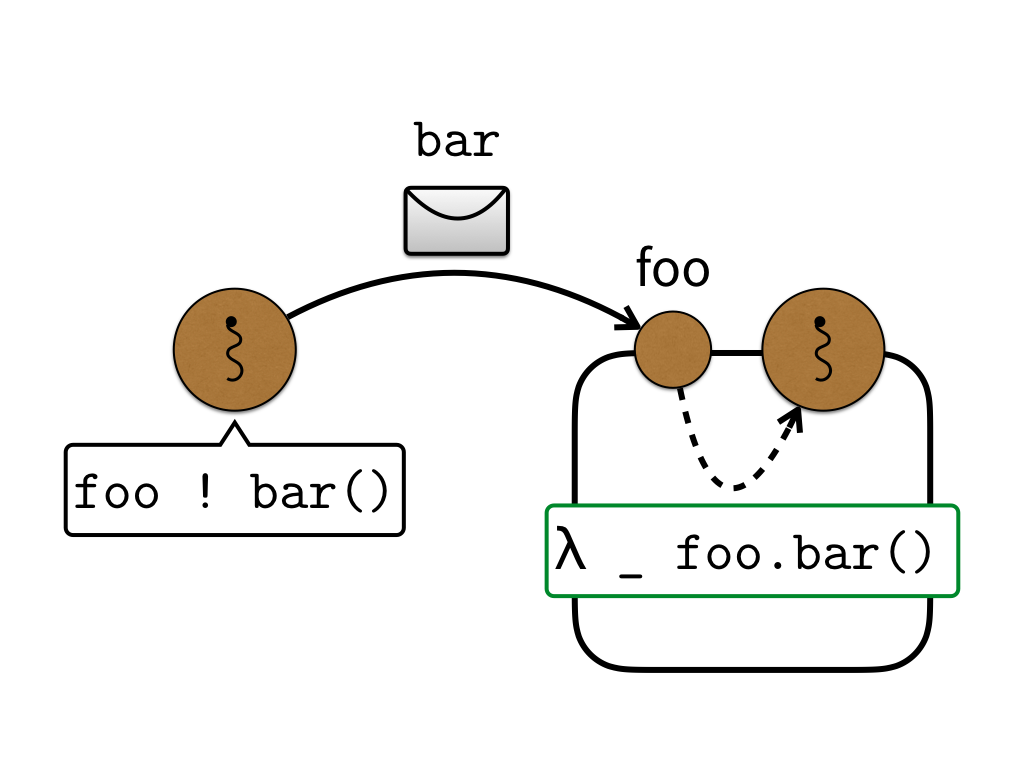}
  \caption{(left) Actor isolation preventing direct manipulation
    of another actor's object. (right) Operations on a reference
    to a passive object belonging to another actor are delegated
    to that actor. We use a \c{.} operator to call methods and
    lookup fields \emph{synchronously}, and a \c{!} operator to
    call methods \emph{asynchronously} (aka \emph{message
      sends}).}
  \FigLabel{illustrate}
\end{figure}

\subsection{Delegation: Motivating Example}
\SecLabel{example}

We motivate breaking isolation in the context of an
object-oriented actor language, with actors serving as the units
of encapsulation, encapsulating zero or more passive objects.
\FigRef{list} shows an Encore program\footnote{We deviate from
  Encore syntax for brevity and omit \textbf{end} keywords in
  favour of using indentation to show block scoping.} with a
linked list in the style of an actor with an asynchronous external
interface. For simplicity we allow asynchronous calls to return
values and omit the details of how this is accomplished (\eg{} by
using futures, promises, or by passing continuations).
We use the \c{.} operator to denote synchronous method calls or
field lookups, and the \c{!} operator to denote asynchronous
method calls.

Clients can interact with the list for example by sending the
message \c{get} with a specified index. With this implementation,
each time \c{get} is called, the corresponding element is
calculated from the head of the list, giving linear time
complexity for each access. Iterating over all the elements
of the list has quadratic time complexity, which is clearly undesirable.

\begin{figure}[t]
  \begin{subfigure}[t]{.45\textwidth}
\begin{code}
class Node[t]
  var next : Node[t]
  var elem : t
  // getters and setters omitted

actor List[t]
  var first : Node[t]
  def getFirst() : Node[t]
    return this.first

  def get(i : int) : t
    var current = this.first
    while i > 0 do
      current = current.next
      i = i - 1
    return current.elem
\end{code}
\Tighten
    \caption{\FigLabel{list}}
\Tighten
\end{subfigure}
\begin{subfigure}[t]{.45\textwidth}
\begin{code}
class Iterator[t]
  var current : Node[t]
  def init(first : Node[t]) : void
    this.current = first

  def getNext() : t
    val elem = this.current.elem
    this.current = this.current.next
    return elem

  def hasNext() : bool
    return this.current != null

actor List[t]
  def getIterator() : Iterator[t]
    val iter = new Iterator[t]
    iter.init(this.first)
    return iter
\end{code}
\Tighten
  \caption{\FigLabel{iterator}}
\Tighten
\end{subfigure}
\caption{(a) A list implemented as an actor. (b) An iterator for
  that list. }
\end{figure}

% \begin{minipage}{\textwidth}
% \begin{code}
% val len = list!length()
% var i = 0
% while i < len do
%   val elem = list!get(i) // Linear time operation
%   ...
% \end{code}
% \end{minipage}

% \noindent
To allow more efficient element access, lists commonly
provide an iterator which holds a
pointer to the current node (\FigRef{iterator}). This allows
constant-time access to the \emph{current} element, and linear
iteration, but also breaks encapsulation by providing direct
access to nodes and elements without going through the list
interface (\cf{} \cite{potanin2013your} for a related discussion of
internal vs. external iterators). Since a list is its own actor,
with this design, list operations are now subject to data-races,
which breaks the single-thread-of-control abstraction.

% A more efficient implementation for this use-case uses an
% \c{Iterator} (\FigRef{iterator}). It stores the current node of an
% ongoing iteration, allowing constant-time lookup of the next
% element:

% \begin{minipage}{\textwidth}
% \begin{code}
% val iter = list!getIterator()
% while iter.hasNext() do
%   val elem = iter.getNext() // Constant time operation
%   ...
% \end{code}
% \end{minipage}

A middle ground providing linear time iteration without data-races can be implemented by
% However, the iterator requires access to the internal state of the
% list actor, meaning that concurrent usages of the list and the
% iterator could lead to data-races, for example removing a link
% while an iterator is accessing it. A simple solution would be to
moving the iterator logic into the list actor, so that
the calls to \c{getNext} and \c{hasNext} are synchronised in the
message queue of the actor. This
% This solves the data-race issue \RED{but}, is
% unsatisfactory for two reasons: First, it limits the number of
% concurrent iterators to one (as there is only one \c{current}
% field), or
requires a more advanced scheme to map different
clients to different concurrent iterators, clutters the
list interface, % and forces the
% list to provide all the operations of the iterator by making it
% part of the list interface, which
creates unnecessary coupling
between \c{List} and \c{Iterator}, and complicates support of
\eg{} several kinds of iterators.

The next section shows how external accesses to an internal object
can be synchronised to preclude data-races, while still avoiding
all of the complications mentioned above.

\begin{figure}[t]
\begin{minipage}[t]{.53\textwidth}
\begin{code}
actor List[t]
  ...
  def getIterator() : B(Iterator[t])
    val iter = new Iterator[t]
    iter.init(this.first)
    return bestow iter
\end{code}
\end{minipage}
\begin{minipage}[t]{.47\textwidth}
\begin{code}
val iter = list!getIterator()
while iter!hasNext() do
  val elem = iter!getNext()
  ...
\end{code}
\end{minipage}
    \Tighten
\caption{\FigLabel{bestow} A list actor returning a bestowed
  iterator, and the code for a client using it}
    \Tighten
\end{figure}

\section{Bestow: Adding Delegation to Encore}
\SecLabel{bestow}

% \newcommand{\Bestow}{\textcolor{blue}{\rm\textbf{bestow}}}

% As the example of the previous section showed, it sometimes makes
% sense for an actor to share some of its internal state, but that
% doing so risks introducing data-races or unwanted dependencies
% between actors and passive objects. In this section we propose a
% solution to this problem that allows leaking private data, while
% still maintaining proper synchronisation.

Encapsulating state behind a synchronisation mechanism allows
reasoning sequentially about operations on that state. The Kappa
type system in Encore will not allow calls to
\c{getIterator()} from \emph{outside} the list actor itself,
defeating the purpose of creating the iterator in the first place.
In Encore, the \c{Iterator} return type of \c{getIterator()} has a
subordinate mode (\cf{} \SecRef{background}), which is a requirement for allowing the direct
reference to the actor's links. Similar to ownership types
\cite{OT}, calls to methods that return subordinate objects are
only allowed on other subordinate objects, \ie{} navigation is
permitted inside each isolation domain, but not across. Notably,
the \c{this} reference of an actor has a subordinate mode, meaning
that an actor's internal type differs from its external.

A call on a non-subordinate receiver thus always denotes a call
that crosses from one synchronisation domain into another. To
support delegation-based synchronisation in Encore, we extend
Kappa with a \emph{bestowed} mode, written as \c{B(T)} when
applied to the type \c{T}. \c{B(T)} denotes an external reference
to an object of type \c{T} which is part of the internal state of
an actor. All of \c{T}'s operations are available through a
reference of type \c{B(T)}, but they will be performed
asynchronously by implicitly delegating the operations to the
actor that owns the object.
% Thus, if \c{def frob(x : T1) : T2} is part of \c{T}, \c{B(T)}
% contains \c{def frob(x : T1) : Fut[T2]} where the return type is
% changed into a future.

For clarity, we introduce a \emph{bestow} operation, which lifts a
type with subordinate mode to an equivalent type whose mode is
instead bestowed. This operation allows a programmer to create
a ``safe remote reference'' to a local passive object, which can be
freely shared---and safely, since it adheres to the delegation model.
We use the term \emph{bestowed reference} to mean a reference to a
\emph{bestowed object}, which in turn denotes an object to which
there are bestowed references. We call the actor encapsulating the
bestowed object its \emph{owner}.

% The goal of actor isolation is to allow sequential reasoning. Breaching
% encapsulation by sharing state is not inherently problematic, as
% long as accesses to shared state are always synchronised with the
% activity of the owner of that state. Based on this observation, we
% propose a mechanism to \emph{bestow activity} on a passive object.
% This makes the passive object appear as an active object, but
% behind the scenes, all operations on the object is performed by
% its owning actor. This allows sharing internal objects freely with
% the outside world, without giving up control on when operations on
% those objects are carried out.

% If \c{y = bestow x}, the message send \c{y!foo()} can be seen as
% syntactic sugar for the operation \c{y.owner!perform(}$\lambda$\c{
%   \_ . x.foo())}, where \c{y.owner} is an implicit field storing a
% reference to the actor that created the bestowed object, and
% \c{perform} is an implicit behavior that tells the actor to
% execute an anonymous function.

\FigRef{bestow} shows the changes needed to the code in
\FigRef{iterator}, as well as the code for a client using the
iterator, to go from isolation-based to delegation-based data-race
freedom. In the list, \c{getIterator()} now returns a bestowed
iterator which is tractable through the \c{B(...)}
type\footnote{If desired, this type change can be implicit through
  view-point adaptation \cite{the-mueller}; Kappa lets us statically
detect when an internal object is crossing its encapsulation boundary.
For clarity, we explicitly annotate bestowed return types here.}, and not
a passive iterator. In the client code, synchronous calls to
\c{hasNext()} and \c{getNext()} become asynchronous message sends
which is reflected by the change from \c{.} to \c{!}. Notably,
messages to the iterator are handled by the list actor and are
and are executed in a serialised fashion, interleaved with normal
messages sent to the list actor.

The granularity of the delegated operations, and thereby the
amount of interleaving allowed, is related to the reach of the
objects which are bestowed. Creating an iterator \emph{inside} the
list and returning a bestowed reference to it allows it to access
multiple nodes of the list atomically. In contrast, creating an
iterator \emph{outside} the list holding bestowed references to
links in the list only gives atomic access to a single link at a
time.
In the case of returning a bestowed reference to an iterator,
entire \c{getNext()} operations are performed without interleaved
activities in the list actor. In the case of returning bestowed
references to individual links, getting the element out of a link,
and obtaining a bestowed reference to its next link are \emph{two}
separate operations which may be interleaved with other operations
on the list (\cf{} \FigRef{fine-grained-iterator}).

\subsection{Bestowed References and Lock-Based Synchronisation}

Encore actors are typed by Kappa's actor mode, which is visible in
the code examples so far. There, the actor mode is applied to
class declarations to denote that instances of these classes are
data-race free because of their asynchronous interface and
encapsulated single thread of control. Replacing ``actor'' with
``locked'' in the code examples up to this point is sound and
preserves data-race freedom of all instances of locked classes by
forcing all method calls to first acquire a per-instance lock.

Notably, the bestowed type and bestow operations in
\FigRef{bestow} are still meaningful if \emph{actor} is changed
for \emph{locked}. At the type-level, a bestowed reference is
oblivious to \emph{how} an isolated object's data-race freedom is
guaranteed. Dynamically, though, a bestowed reference into a
capsule protected by a lock will implicitly acquire the lock
associated with that capsule.

\subsection{Differences Between Locks and Actors}
\SecLabel{LocksVsActors}

The small changes needed in Kappa to switch between using locks
and actors as a means of protecting access to a group of objects
does not reflect the differences between the two concepts. In
lock-based programs, data is ``dead'' and only lives when visited
by a thread of control. In actor-based programs, actors are alive,
only act upon themselves, and only upon so choosing. The
difference is push vs. pull---whether activity is pushed upon an
object by a thread, or whether activity happens as a result of the
actor pulling a message from its message queue. In lock-based
programs, a thread can be denied entrance into an object by
another thread holding the appropriate lock. In actor-based
programs, an actor may never ask for the next message (for example
because it is stuck in an infinite loop) causing another actor's
request to be effectively ignored.

In an actor-based setting, the owner of a bestowed object will
perform all operations on that object. Thus, while the owner will
technically never block on an access to an object that it owns, it
may instead end up performing operations on behalf of other
actors. In some scenarios, this may make the owner of a bestowed
object an unintentional bottleneck in a system, especially if
delegated operations are long-running, the owner has
unproportionally many bestowed objects, or it performs critical
operations unrelated to the bestowed object which are now
interleaved with delegated tasks. In a lock-based setting, all
threads operating on a shared resource take part in performing
those operations. Thus, if operations are spread out in time, it
is possible that contention is never witnessed in the system and
that operations unrelated to the bestowed object are never
delayed.

\subsection{Towards Actor-Style Delegation-Locks}

We note that it is possible to make a bestow-based interpretation
of locks by replacing delegation with transfer of ownership:
instead of an actor lifting an operation into a closure and
passing it to the owner to be performed, the actor simply
transfers ownership of the bestowed object to itself and
subsequently performs the operation without any delegation.

Because there are no topological restrictions on bestowed
references, transfer of ownership of a bestowed object $o$ is
possible unless $o$ is currently being used (equivalent to another
thread holding a lock).
For example, if an actor which owns some collection data structure
is currently inactive, another actor with a bestowed reference to
this collection could transfer ownership of the collection to
itself and operate on it, rather than waking the owning actor to
delegate the operation.

If transfer is not possible, the actor
attempting the transfer could block (making accesses to bestowed
references equivalent to accessing references with locked mode) or
fall-back to delegation to avoid blocking. This is similar to
\emph{queue delegation locks} \cite{QDL} which handle contention by
delegating blocking operations to the thread currently holding the
lock. This allows operations that would otherwise block to return
immediately with a future value which will eventually be fulfilled
by some other thread.

Allowing objects' owners to change prevents giving the owner
special access, and thus forces all references to a bestowed
object to be bestowed. Accesses must then branch on whether the
object is local, and if not, whether or not its ownership can be
transferred. For the example above, whoever is accessing the
collection must first check if they are the current owner of the
collection, in which case the operation can proceed, and otherwise
decide if ownership transfer is possible or if the operation
should be delegated.
\SecRef{transfer} gives the details on the semantics of
transferable bestowed objects.

\SecLabel{load-balancing}
Encore uses a work-stealing based load-balancing scheme that moves
actors across cores. In many ways, allowing transfer of ownership
of bestowed objects is a form of load-balancing for passive
objects. Just like load-balancing of actors is governed by
carefully tuned heuristics, how and when to transfer ownership of
a bestowed object is not clear-cut. In many cases, keeping objects
local to a core may have locality benefits which favours ownership
transfer for actors on the same core, but delegation
otherwise %\footnote{This is also interesting to ponder in a
%  distributed setting.}.
%
Also, the cost of creating and passing
closures may be non-negligible, which may favour keeping ownership
among frequent accessors, etc. Run-time optimisation of ownership
transfer for bestowed references is an interesting direction for
future work.

\section{Atomic: Atomicity and Grouping of Operations on Bestowed References}
\SecLabel{atomic}

So far, we have discussed data-race freedom. Data-race freedom is
important, but not always enough achieve atomicity, \ie{} the
ability to make a set of changes appear atomic in a system. For
example, a data-race free counter in Kappa will correctly handle
concurrent increments and decrements, but is not enough to support
swapping the values of two counters in the presence of concurrent
operations. The root cause is that Kappa's protection is per-object
(notably including its transitive closure of sub-objects), and
exchanging values involves more than one object where neither
object is nested inside the other.

Additionally, the protection from data-races offered by Kappa only
applies to a single top-level operation\footnote{Kappa can
  overcome this limitation through explicit lock-taking, which is
  only supported on locked objects, not actors \cite{castegren16}.}. For
example, operating on a \c{locked} file to first \emph{open} and
subsequently \emph{read}, one cannot exclude the possibility of a
separate thread \emph{closing} the file in-between. The same holds
if the file is an actor: messages are serialised but any messages
may be interleaved, meaning that $[\c{open}, \c{close}, \c{read}]$
is a valid mailbox. This problem is exacerbated by more
fine-grained operations being added to an interface through
bestowed references.

To support atomicity of groups of operations over one or more
objects we extend Encore with an atomic block construct. In
its simplest form, it supports operations on a single reference
whose mode is actor, locked, or bestowed and coalesces operations
on the reference making them atomic in the system.

\begin{figure}[t]
\begin{subfigure}[t]{.55\textwidth}
\begin{minipage}[t]{.55\textwidth}
\begin{code}
class Iterator[t]
  var current : B(Node[t])
  def getNext() : t
    val elem = this.current ! elem()
    // Possible interleaving of other messages
    this.current = this.current ! next()
    return elem
\end{code}
\end{minipage}
\caption{\FigLabel{fine-grained-iterator}}
\end{subfigure}
\begin{subfigure}[t]{.47\textwidth}
\begin{minipage}[t]{.47\textwidth}
\begin{code}
class Iterator[t]
  var current : B(Node[t])
  def getNext() : t
    atomic c <- this.current
      val elem = c ! elem()
      // No possible interleaving
      this.current = c ! next()
      return elem
\end{code}
\end{minipage}
\caption{\FigLabel{coarse-grained-iterator}}
\end{subfigure}
    \Tighten
    \caption{Fine-grained (a) and coarse-grained (b) concurrency control.}
    \FigLabel{atomic}
    \Tighten
\end{figure}

As a concrete, simple example, \FigRef{coarse-grained-iterator}
shows an iterator with a bestowed reference to its current node,
and which wraps its operations in an atomic block, preventing
any interleaving between the two message sends, thus regaining the
same interleaving guarantees as the iterator in \FigRef{bestow}.
The atomic block thereby allows a client to define new operations by
composing smaller ones and giving a \emph{use-site} definition of
where atomicity is useful, as opposed to declaration-site. For
example, merging \c{elem} and \c{next} into a single operation in the \c{Node} class
voids the need for an atomic block in
\FigRef{coarse-grained-iterator}, but in the general case, it is
unreasonable to expect declarations to cater to all possible use
cases.

The simplest case, a single atomic block of messages whose results
are ignored (or \c{void}), can be implemented efficiently through
\emph{coalescing} messages to the receiver and delivering them as
a single ``big message''\footnote{In Encore, this may additionally
  have garbage collection benefits if messages contain overlapping
  state.}. \FigRef{atomic-semantics} (left) shows how an atomic
block containing two message sends is translated into a single
message, containing instructions for performing the two operations
in sequence. This avoids interleaving from other actors, but does
not support multi-actor atomicity.
Similarly, for the case of operations on an object guarded by a
lock, an atomic block acts as an explicit locking
operation---similar to a \c{synchronized} block in Java---voiding
the need for an individual lock--release for each operation in the
block (\cf{} \cite{castegren16}).

\subsection{Grouping Semantics and Atomicity}
\SecLabel{grouping-semantics}

Coalescing messages to avoid interleaving is simple and efficient,
but also limited. For example, a client of the iterator from \FigRef{bestow}
may want to atomically perform operations on all the elements of
list until a certain element is found. However, a coalescing
semantics prevents the client from reacting to results from the
iterator except between discrete atomic blocks. Furthermore, it
imposes some upper limit on the messages from the client to the
list. Last, it will delay the time when the client can start
operating on the elements until the right element is found, making
it impossible to implement parallel pipelines. Moreover,
coalescing semantics is not strong enough to support a notion of
atomicity across multiple actors.

Atomicity across actors can be made possible by delaying the
processing of messages not originating from within the atomic
block until after a particular point. In its simplest form, the
actor executing the atomic block will be able to reason about the
receiving actor throughout the atomic block enabling \eg{} pre-
and post condition reasoning. By wrapping several atomic blocks,
this extends to multiple actors.

To this end, we explore a richer semantics for atomic blocks which
involves the creation of a \emph{private} mailbox, which is only
known by the actor running the atomic block and the receiving
actor. Private channels can be found \eg{} in Pi calculus
\cite{milner1999communicating} and CSP
\cite{hoare1978communicating}, but in our implementation we restrict them to follow
a block-scoped, structured programming approach. Under the private
mailbox semantics, at the start of an atomic block, the current
actor creates a new mailbox, technically a FIFO queue, and passes
one end (for receiving) to the target actor in a normal message
and keeps the other for sending. Upon receipt of this message, the
target actor replaces its normal, public mailbox with the private
mailbox, and from this point continues to process only messages
from the latter. Since the private mailbox is only known by the
sender and receiver, no other messages may interleave their
communication. At the end of the atomic block, the target actor is
sent a new message to restore its original mailbox, where any other communication
has been buffered, and will now be processed in FIFO order. (See
\SecRef{money-transfer} for an example that can be encoded with
private mailboxes, but not with coalescing.)

In languages which allow actors to use pattern matching to
selectively pick messages from its message queue (\eg{} Erlang
\cite{erlang}), this kind of ``private conversation'' between two
actors can be encoded. However, this requires that the receiving
actor is implemented with support for atomically processing
exactly the messages that the client wanted to send. By using a
private queue created by the client, the client can create new
atomic operations by composing the operations that a receiving
actor provides, without explicit support for this in the receiving
actor.

The atomic block notably turns actor-based programs into more
structured programs, because the atomic blocks are unidirectional:
the only way for the list actor to communicate with a client during the execution of
an atomic block in the client is through futures---not message sends, as
they will not be processed before the atomic block is finished.
Furthermore, only tree-shaped communication is possible inside
atomic blocks. If a client of the list has entered an atomic block
targeting the iterator, and then enters a nested block targeting some
other actor $A$, there is no way for $A$ and the list actor to communicate
directly\footnote{However, the list actor may return a value to the client which the client
  subsequently passes to $A$, and similar for $A$.}.

\begin{figure}[t]
  \centering
  \includegraphics[width=.4\textwidth]{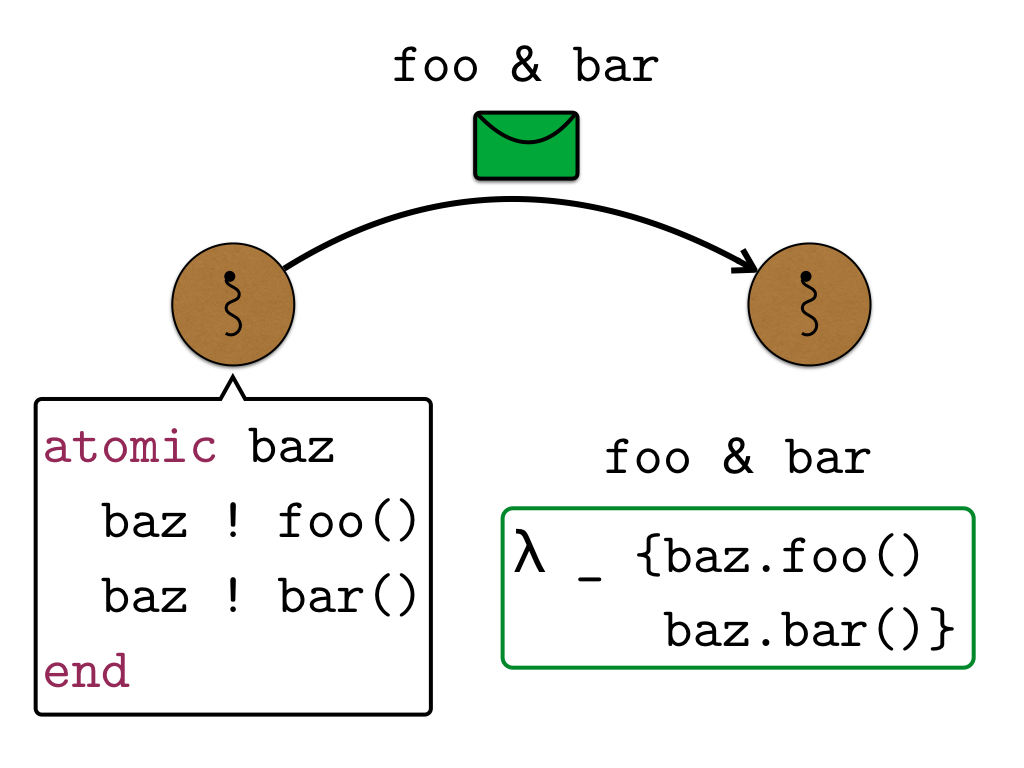}
  \qquad
  \qquad
  \includegraphics[width=.4\textwidth]{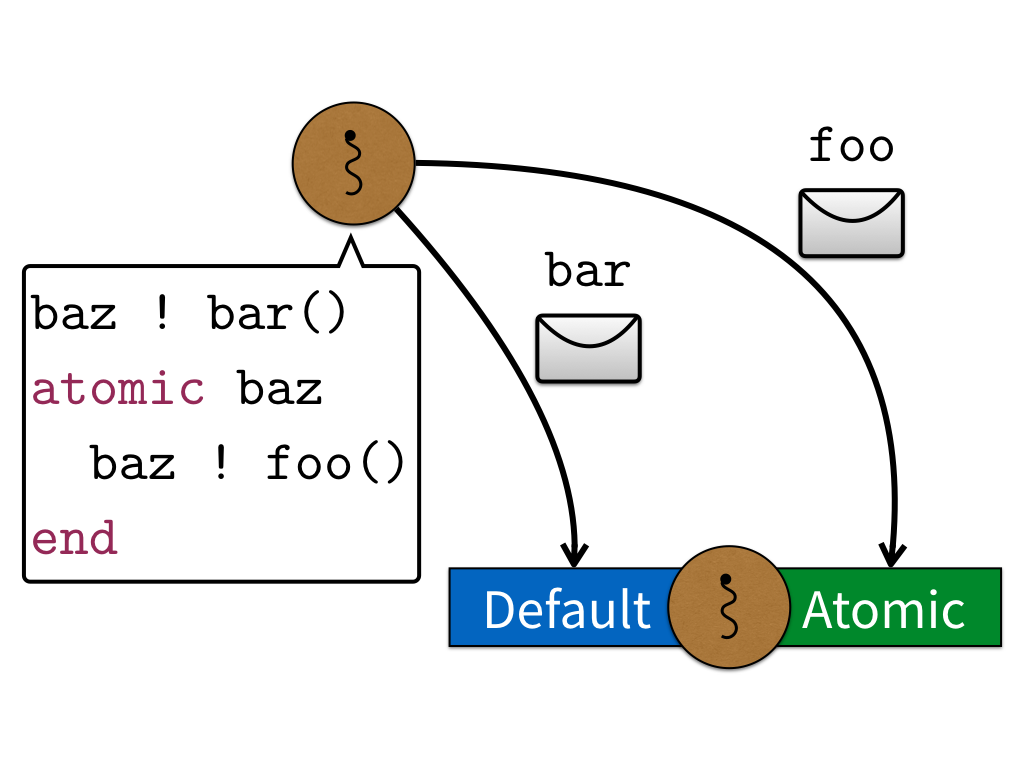}
  \caption{(left) Coalescing semantics. (right) Private mailbox.}
  \FigLabel{atomic-semantics}
\end{figure}

\subsection{Tractability of Atomic Operations}

One reason why lock-based programming is hard in mainstream
programming is because there is no way to express what a lock
governs in a program. Kappa addresses this by keeping track of
synchronisation and encapsulation through modes in types. Imagine
an actor $A$ holding a reference the list actor in its field $f$. In
its method $m_1$, $A$ enters into an atomic block targeting the list actor.
$A$ subsequently calls its method $m_2$ which performs \c{this.f !
  getFirst()}. There is nothing in $f$'s type that expresses the
current situation where communication with the list actor is uninterruptible.
Should \c{this.f ! getFirst()} be considered part of the atomic block
or not, \ie{} is atomicity \emph{actor-wide}? If yes, then the
tractability of Kappa is significantly weakened. If no, we deviate
from the standard behaviour of locks and must additionally solve a
problem with tracking bestowing references. A simple approach in
the latter case is to tie operations to the \emph{lexical scope}
of the atomic block. This means that \c{this.f ! getFirst()} will not
be processed by its receiver until after the atomic block is
finished. A hybrid approach is to introduce an additional
\emph{handle} to the subject of an atomic block, whose type
reflects its ``atomic status''. This handle can be passed to where
atomic operations should be performed, allowing \c{handle ! getFirst()} to
take part in the atomic block. This is the route Encore follows;
inside the block \c{atomic x do ...} the type \c{T} of \c{x} is
changed to \c{A(T)}, where the new qualifier \c{A()} expresses
that:
\begin{enumerate}
\item Operations will not be interleaved with operations from other
  actors.
\item The reference is stack-bound (or \emph{borrowed}) meaning that for
  any given scope where the reference occurs, no aliases can be
  created that survives that scope.
\item The reference is local to the current actor meaning it may
  not be passed to another actor, either as an element or by
  returning it.
\end{enumerate}

In Encore, for the message \c{getFirst()} in $m_2$ to be be part of
the atomic block in $m_1$, $m_2$ must be passed the correct
\c{A()} handle and \c{this.f ! getFirst()} replaced by \c{handle !
  getFirst()}.

% . The situation sketched in
% \SecRef{example}, where a client wants to access two adjacent
% nodes in the list actor without interleaving operations from other
% clients is easily resolved by wrapping the two calls to \c{get}
% (or \c{getNext}, if the iterator is used) inside an \c{atomic}
% block. This will batch the messages and ensure that they are
% processed back to back:

% \begin{minipage}[t]{.45\textwidth}
% \begin{code}
% atomic it <- list ! getIterator()
%   val e1 <- it.getNext()
%   val e2 <- it.getNext()
% \end{code}
% \end{minipage}
% %
% \begin{minipage}[t]{.05\textwidth}
% \vspace*{3 mm}
% $\implies$\quad
% \end{minipage}
% %
% \begin{minipage}[t]{.45\textwidth}
% \begin{code}
% (e1, e2) =
%   list ! ?$\lambda$? this .
%     {val it = this.getIterator();
%      val e1 = it.getNext();
%      val e2 = it.getNext();
%      return (e1, e2)}
% \end{code}
% \end{minipage}

\section{Formalising ``Vanilla'' Bestow and Atomic}
\SecLabel{formalism}
\SecLabel{formal1}

To explain bestow and atomic we use a simple lambda
calculus with actors and passive objects. For brevity, we only
discuss bestow and atomic under one concurrency model, namely
actors. We chose actors for two reasons: they have the most
compelling design space, and it more closely matches our
implementation in Encore. In this section, we formalise the bestow
as presented in \SecRef{bestow} together with the coalescing
atomic operations from \SecRef{atomic}.
In subsequent sections, we also formalise two variants of the
actor-based semantics for bestow and atomic: passive
objects with transferable ownership and private mailboxes.

The purpose of these formalisations is to flesh out the details of
bestow and atomic,
and we abstract away most details that are unimportant for describing
the behavior of these constructs. For example, we leave out
classes and actor interfaces and simply allow arbitrary operations
on values. This makes the formalisation simple and fairly unsurprising.
By disallowing sharing of (non-bestowed) passive
objects, we show that our language is free from data-races
(\CF{meta}). The calculus has been mechanised and proven sound in
Coq (\CF{coq}).

The syntax of our calculus is shown in \FigRef{syntax}. An
expression $e$ is a variable $x$, a function application $e$ $e'$
or a message send $e\c{!}v$. Messages are sent as anonymous
functions, which are executed by the receiving actor.
We abstract updates to passive objects as $e.\c{mutate()}$, which
has no actual effect in the formalism, but is reasoned about in
\SecRef{meta}. A new object or actor is created with \c{new}
$\tau$ and a passive object can be bestowed by the current actor
with \c{bestow} $e$.
Since this version of the semantics uses coalescing semantics
for atomic blocks, we don't yet need a special \c{atomic}
construct in the formalism. Instead we model atomic interactions by
composing operations as in \FigRef{atomic-semantics} (left).

\begin{figure}[t]
\begin{tabular}{rcl}
$e$ & $::=$ &  $x$
           $|$ $e$ $e$
           $|$ $e\c{!}v$
           $|$ $e.$\c{mutate()}
           $|$ \c{new} $\tau$
           $|$ \c{bestow} $e$
           $|$ $v$ \\
$v$ & $::=$ &  $\lambda x : \tau . e$
           $|$ $()$
           $|$ $\mathit{id}$
           $|$ $\iota$
           $|$ $\iota_{\mathit{id}}$ \\
\end{tabular}
\quad
\begin{tabular}{rcl}
$\tau$ & $::=$ & $\alpha$
           $|$ $\mathsf{p}$
           $|$ $\tau \rightarrow \tau$
           $|$ \texttt{Unit} \\
$\alpha$ & $::=$ & $\mathsf{c}$
           $|$ $\mathbf{B}(\mathsf{p})$ \\
\end{tabular}
\caption{\FigLabel{syntax} The syntax of a simple lambda calculus
  with actors and a \c{bestow} operation.}
\end{figure}

Statically, values are anonymous functions or the unit value $()$.
Dynamically, $\mathit{id}$ is the identifier of an actor, $\iota$ is the
memory location of a passive object, and $\iota_{\mathit{id}}$ is
a passive object $\iota$ bestowed by the actor $\mathit{id}$.
A type is an active type $\alpha$, a passive type $\mathsf{p}$, a function
type $\tau \rightarrow \tau$, or the \texttt{Unit} type. An active type
is either an actor type $\mathsf{c}$ or a bestowed type $\mathbf{B}(\mathsf{p})$. Note that for simplicity,
$\mathsf{p}$ and $\mathsf{c}$ are not meta-syntactic variables; every passive object
has type $\mathsf{p}$, every actor has type $\mathsf{c}$, and every bestowed
object has type $\mathbf{B}(\mathsf{p})$.

\begin{figure}[th]
\drules[e]
  {$\Gamma \vdash e : \tau$}
  {Expressions}
  {var
  ,apply
  ,newXXpassive
  ,newXXactor
  ,mutate
  ,bestow
  ,send
  ,fn
  ,unit
  ,loc
  ,id
  ,bestowed
  }
  \caption{\FigLabel{static} Static semantics. $\Gamma$ maps
    variables to types. $\Gamma_{\alpha}$ contains only the active
    types $\alpha$ of $\Gamma$.}
\end{figure}

\subsection{Static Semantics}

The typing rules for our formal language can be found in
\FigRef{static}. The typing context $\Gamma$ maps variables to types.
The ``normal'' lambda calculus rules \ARN{e-var} and \ARN{e-apply}
are straightforward. The \c{new} keyword can create new passive
objects or actors \RN{e-new-*}. Passive objects may be mutated
\RN{e-mutate}, and may be bestowed activity \RN{e-bestow}.

Message sends are modelled by sending anonymous functions which are
run by the receiver \RN{e-send}. The receiver must be of active
type (\ie{} be an actor or a bestowed object), and the argument of
the anonymous function must be of passive type $\mathsf{p}$ (this can be
thought of as the \c{this} of the receiver). Finally, all free
variables in the body of the message must have active type to make
sure that passive objects are not leaked from their owning actors.
This is captured by $\Gamma_{\alpha}$ which contains only the
active mappings $\_ : \alpha$ of $\Gamma$. Dynamically, the body may
not contain passive objects $\iota$.
Typing values is straightforward.
% (\ARN{eval-fn}, \ARN{eval-unit}, \ARN{eval-loc}, \ARN{eval-id},
% \ARN{eval-bestowed}).

\subsection{Dynamic Semantics}
\SecLabel{dyn-semantics}

\begin{figure}[t]
\drules[eval]
  {$H \overset{id}{\hookrightarrow} H'$}
  {Evaluation}
  {actorXXmsg
  ,actorXXrun
  }

\drules[eval]
  {$\mathit{id} \vdash \langle H, e \rangle \hookrightarrow \langle H', e' \rangle$}
  {Evaluation of expressions}
  {sendXXactor
  ,sendXXbestowed
  ,apply
  ,mutate
  ,bestow
  ,newXXpassive
  ,newXXactor
  ,context}

  $E[\bullet] ::= \bullet~e ~|~ v~\bullet ~|~ \bullet\c{!}v ~|~
  \bullet.\c{mutate()} ~|~$\c{bestow}$~\bullet$
\caption{\FigLabel{dynamic} Dynamic semantics.}
\end{figure}

\FigRef{dynamic} shows the small-step operational semantics
for our language.
A running program is a heap $H$, which maps actor identifiers $id$
to actors $(\iota, L, Q, e)$, where $\iota$ is the \c{this} of the
actor, $L$ is the local heap of the actor (a set containing the
passive objects created by the actor), $Q$ is the message queue (a
list of lambdas to be run), and $e$ is the current expression
being evaluated.
The step relation $\overset{id}{\hookrightarrow}$ is indexed by
the actor $id$ that is currently scheduled.

An actor whose current expression is a value may pop a message
from its message queue and apply it to its \c{this}
\RN{eval-actor-msg}. Any actor in $H$ may step its current
expression, possibly also causing some effect on the heap
\RN{eval-actor-run}. The relation
$\mathit{id} \vdash \langle H, e \rangle \hookrightarrow \langle
H', e' \rangle$ denotes actor $\mathit{id}$ evaluating expression
$e$ one step in heap $H$, resulting in a new heap and expression
$e'$ and $H'$.

Sending a lambda to an actor prepends this lambda to the
receiver's message queue and results in the unit value
\RN{eval-send-actor}. Sending a lambda $v$ to a bestowed value
instead prepends a new lambda to the queue of the actor that
bestowed it, which simply applies $v$ to the underlying passive
object \RN{eval-send-bestowed}.

Function application replaces all occurrences of the parameter $x$
in its body by the argument $v$ \RN{eval-apply}. Mutation is a
no-op in practice \RN{eval-mutate}. Bestowing a passive value
$\iota$ in actor $\mathit{id}$ creates the bestowed value
$\iota_{\textit{id}}$ \RN{eval-bestow}.

Creating a new object in actor $\mathit{id}$ adds a fresh location
$\iota'$ to the set of the actors passive objects $L$ and results
in this value \RN{eval-new-passive}. Creating a new actor adds a
new actor with a fresh identifier to the heap. Its local heap
contains only the fresh \c{this}, its queue is empty, and its
current expression is the unit value \RN{eval-new-actor}.

We handle evaluation order by using an evaluation context $E$
\RN{eval-context}.

\begin{figure}[th]
  \centering
\drules[wf]
  {$\vdash H \qquad H \vdash (\iota, L, Q, e) \qquad H \vdash Q$}
  {Well-formedness}
  {heap
  ,actor
  ,queueXXmessage
  ,queueXXempty
  }

  \caption{\label{fig:wf} Well-formedness rules. $\mathcal{LH}$
    gets the local heap from an actor:
    $\mathcal{LH}((\iota, L, Q, e)) = L$}
\end{figure}

\subsection{Well-formedness}

\FigRef{wf} shows our well-formedness rules. A heap $H$ is well-formed if all its actors are well-formed with
respect to $H$, and the local heaps $L_i$ and $L_j$ of any two
different actors are disjoint \RN{wf-heap}. We use
$\mathcal{LH}(H(id))$ to denote the local heap of actor $id$.
An actor is well-formed if its \c{this} is in its local heap $L$
and its message queue $Q$ is well-formed. The current expression
$e$ must be typable in the empty environment, and all passive
objects $\iota$ that are subexpressions of $e$ must be in the
local heap $L$. Similarly, all actor identifiers in $e$ must be
actors in the system, and all bestowed objects must belong to the
local heap of the actor that bestowed it \RN{wf-actor}.

A message queue is well-formed if all its messages are well-formed
\RN{wf-queue-*}. A message is well-formed if it is a well-formed
anonymous function taking a passive argument, and has a body $e$
with the same restrictions on values as the current expression in
an actor.

\subsection{Meta Theory}
\SecLabel{meta}

We prove soundness of our language by proving progress and
preservation in the standard fashion:

\begin{quote}
  \textbf{Progress}: In a well-formed heap $H$, each actor can
  either be evaluated one step, or has an empty message queue and
  a fully reduced expression:
\[
  \forall id \in \mathbf{dom}(H).
  \vdash H \implies
    (\exists H' ~.~ H \overset{id}{\hookrightarrow} H')
    ~\lor~
    (H(id) = (\iota, L, \epsilon, v))
\]
\end{quote}

\begin{quote}
  \textbf{Preservation}: Evaluation preserves well-formedness of heaps:
$
  \vdash H ~ \land ~ H \overset{id}{\hookrightarrow} H' \implies
    \vdash H'
$
\end{quote}

\noindent
Both properties can be proven to hold with straightforward
induction.

The main property that we are interested in for our language is
data-race freedom. As we don't have any actual effects on passive
objects, we show this by proving that if an actor is about to
execute $\iota.$\c{mutate()}, no other actor will be about to
execute \c{mutate} on the same object:

\begin{quote}
  \textbf{Data-race freedom}: In a well-formed system, for any two
  possible reductions that involve two different actors each
  mutating a passive object, the respective passive objects can
  never be the same.
\[
\left(
\begin{array}{c}
id_1 \neq id_2 \\
\land~ H(id_1) = (\iota_1, L_1, Q_1, \iota.\texttt{mutate}())\\
\land~ H(id_2) = (\iota_2, L_2, Q_2, \iota'.\texttt{mutate}())
\end{array}
\right)
\implies \iota \neq \iota'
\]
\end{quote}
\noindent
This property is simple to prove using two observations on what
makes a well-formed heap:

\Pad
\begin{compactenum}
\item An actor will only ever access passive objects that are in
  its local heap \RN{wf-actor}.
\item The local heaps of all actors are disjoint \RN{wf-heap}.
\end{compactenum}
\Pad

\noindent
The key to showing preservation of the first property is in the
premise of rule \ARN{e-send} which states that all free variables
and values must have an active type
($\Gamma_\alpha, x : \mathsf{p} \vdash e' : \tau'$ and $\not\exists \iota ~.~ \iota \in e'$). This prevents
sending passive objects between actors without bestowing them
first. Sending a message to a bestowed object will always relay
it to the actor that owns the underlying passive object
(by the premise of \ARN{wf-actor}: $\forall \iota_{id} \in e ~.~ \iota \in \mathcal{LH}(H(id))$).
Preservation of the second property is simple to show since
local heaps grow monotonically, and are only ever extended with
fresh locations \RN{eval-new-passive}.

Having made these observations, it is trivial to see that an actor
in a well-formed heap $H$ that is about to execute
$\iota.$\c{mutate()} must have $\iota$ in its own local heap. If
another actor is about to execute $\iota'.$\c{mutate()}, $\iota'$
must be in the local heap of this actor. As the local heaps are
disjoint, $\iota$ and $\iota'$ must be different. Since
well-formedness of heaps are preserved by evaluation, all programs
are free from data-races.

Finally, we show that atomic blocks indeed provide atomicity. For
this version of the semantics, we do not state this property
formally;
since atomic blocks are implemented as ``big messages'', and
messages are always run to completion before another message is
handled, it follows trivially that atomic operations run without
any interleaving in this semantics.

\subsection{Mechanisation}
\SecLabel{coq}

The formalism presented in this section has been fully mechanised
in Coq, including proofs of progress and preservation
\cite{bestow-repo}. The mechanised version is true to the paper
version, modulo uninteresting differences in representation (for
example, the heap is a list of actors indexed by actor $id$s,
rather than a map from $id$s to actors as in
\SecRef{dyn-semantics}). It also includes machinery for generating
fresh values.

The whole definition of the semantics is $\approx$450 lines of Coq. The
proofs are $\approx$2100 lines, including $\approx$340 lines of auxiliary lemmas
about list operations and $\approx$250 lines of tactics specific to this
formalism, including automated case analysis for heap operations.
The proofs also make use of the LibTactics library
\cite{libtactics}, as well as the \c{crush} tactic \cite{cpdt}.
There is still some repetition between similar cases in some of
the proofs, which could be broken out to further reduce the size
of the proof.

\section{Formalising Private Mailboxes and Bestow with Ownership Transfer}
\SecLabel{formal2}

This section explores two variations on the semantics presented in
\SecRef{formalism}; bestowed objects that may change owner, and
atomic message passing using private message queues. Both of these
variations have also been fully mechanised and proven sound in
Coq \cite{bestow-repo}.

\subsection{Transferring Ownership of Bestowed Objects}
\SecLabel{transfer}

\begin{figure}[t]
\begin{tabular}{rcl}
$e$ & $::=$ &  $x$
           $|$ $e$ $e$
           $|$ $e\c{!}v$
           $|$ $e.$\c{mutate()}
           $|$ \c{new} $\tau$
           $|$ $v$ \\
$v$ & $::=$ &  $\lambda x : \tau . e$
           $|$ $()$
           $|$ $\mathit{id}$
           $|$ $\iota$
           $|$ $\iota_*$ \\
\end{tabular}
\quad
\begin{tabular}{rcl}
$\tau$ & $::=$ & $\alpha$
           $|$ $\mathsf{p}$
           $|$ $\tau \rightarrow \tau$
           $|$ \texttt{Unit} \\
$\alpha$ & $::=$ & $\mathsf{c}$
           $|$ $\mathbf{T}(\mathsf{p})$ \\
\end{tabular}
\caption{\FigLabel{transfer-syntax} The syntax of a simple lambda calculus
  with actors and transferable objects.}
\end{figure}

\begin{figure}[t]
\drules[e]
  {$\Gamma \vdash e : \tau$}
  {Expressions}
  {newXXtrans
  ,trans
  ,sendXXtrans
  }
  \caption{\FigLabel{transfer-static} New and changed static rules
    for dealing with transferable objects.}
\end{figure}

\begin{figure}[t]
\drules[eval]
  {$\mathit{id} \vdash \langle \mathcal{O}, H, e \rangle \hookrightarrow \langle \mathcal{O}', H', e' \rangle$}
  {Evaluation of expressions}
  {newXXtrans
  ,sendXXtransXXrun
  ,sendXXtransXXdelegate
  }

\drules[eval]
  {$\langle \mathcal{O}; H \rangle \overset{id}{\hookrightarrow} \langle \mathcal{O}'; H' \rangle$}
  {Evaluation}
  {actorXXtrans
  }

\caption{\FigLabel{transfer-dynamic} New and changed dynamic rules for dealing with transferable objects.}
\end{figure}

\begin{figure}[th]
  \centering
\drules[wf]
  {$H \vdash \mathcal{O} \qquad \mathcal{O}; H \vdash (\iota, L, Q, e) \qquad \mathcal{O}; H \vdash Q$}
  {Well-formedness}
  {owners
  ,actorXXtrans
  ,queueXXmessageXXtrans
  }

  \caption{\FigLabel{trans-wf} New and changed well-formedness
    rules for dealing with transferable objects.}
\end{figure}

So far, bestowed objects have always been passive objects owned by
some actor, and the owner of these objects never changes. This
means that once an object has been bestowed and shared between
actors, messages to it will always be relayed to the same actor.
If an actor bestows many objects, the (implicit) contention on
this actor will be high. If ownership of a bestowed object could
be transferred, some of this contention could be alleviated.

\FigRef{transfer-syntax} shows the syntax of this language
variation. A transferable object has type
$\mathbf{T}(\mathsf{p})$. Since such an object may move between
actors, we cannot allow this object to have aliases of the
non-transferable type $\mathsf{p}$. Here, we solve this by
requiring the object to be created as transferable with \c{new}
$\mathbf{T}(\mathsf{p})$, which voids the need for a \c{bestow}
operation. Dynamically, a transferable object has the value
$\iota_*$. An actor which evaluates $\iota_*\c{!}v$ when it is the
owner of $\iota_*$ will perform $v~\iota$ synchronously. Other
actors will relay to the current owner.
Note that transferring ownership of an object does not prevent the
previous owner from retaining a reference to the transferred
object. The only difference is that messages sent to the
transferred object will now be relayed to the new owner instead.

\FigRef{transfer-static} shows the new and changed rules of the
static semantics. There are two new rules for typing transferable
objects (\ARN{e-new-trans}, \ARN{e-trans}). A subtle change is also
needed in the send rule: the body of the message sent must have
type \c{Unit} to preserve the type of the expression when the
message is run synchronously \RN{e-send-trans}. This is without
loss of generality since actors cannot do anything with the result
of a message send.

\FigRef{transfer-dynamic} shows the new and changed rules of the
dynamic semantics. To track the current owner of a transferable
object, we extend the configuration with a map $\mathcal{O}$ from
locations $\iota$ to actors $\iota$. Creating a new transferable
object adds its creator to the owner map \RN{eval-new-trans}. When
evaluating $\iota_*\c{!}v$, the owner of $\iota_*$ will be looked
up in $\mathcal{O}$ and the message will either be run
synchronously \RN{eval-send-trans-run} or be relayed to the
current owner \RN{eval-send-trans-delegate}. In
\SecRef{formalism}, the ``unpacking'' of the bestowed object
happens upon delegation (\cf{} \ARN{eval-send-bestowed} in
\FigRef{dynamic}). Here, the owner may have changed when the
message arrives, so the message we send contains the original
message send $\iota_*\c{!}v$ to allow further delegation if
needed.

We allow ownership transfer when the current owner is not running
a message \RN{eval-actor-trans}. If ownership could be transferred
in the middle of an actor's behavior, there could be data-races.
When ownership of an object is transferred, its location $\iota$
is also moved between the local heaps of the actors involved in
the transfer. Note that there may still be delegated messages in
the queue of the old owner, and that these will be forwarded to
the new owner when they reach the head of the queue. Ownership
transfer is non-deterministic to simulate a run-time or scheduler
that does load-balancing between actor behaviors.

Finally, \FigRef{trans-wf} shows the new and changed
well-formedness rules. An owner map $\mathcal{O}$ is well-formed
if for all mappings $\iota \mapsto id$, $id$ is an actor which has
$\iota$ in its local heap \RN{wf-owners}. For a well-formed actor,
its \c{this} must not be in $\mathcal{O}$ (\ie{} not be
transferable, since \c{this} of an actor never changes owner), and
for all transferable values in its current expression, there must
be some mapping for that location in $\mathcal{O}$
\RN{wf-actor-trans}. Similar rules hold for messages, but
additionally their locations $\iota$ may not have transferable
ownership \mbox{\RN{wf-queue-message-trans}}, as this would mean
that their ownership could be transferred before the message has
reached the head of the queue.

The adaptation of the meta-theoretic properties is
straightforward, and the same soundness properties hold for this
variation of the formalism. The mechanised version of the
semantics are $\approx$470 lines of Coq, and the proofs are $\approx$2600 lines
with the same $\approx$600 lines of list lemmas and tactics as for the
core calculus \cite{bestow-repo}.

\subsubsection{Design Considerations for Transferable Objects}

For simplicity, this variation does not include the bestowed
references from the formalism in \SecRef{formalism}. It is
possible to combine both flavors of bestowed references in the
same system as long as care is taken to ensure that an object is
never both bestowed and transferable. The only place where such an
object could be created is when a transferable object is being
operated on synchronously (\cf{} \ARN{eval-send-trans-run} in \FigRef{transfer-dynamic}). This
could for example be achieved by introducing a type annotation
which disallows bestowing, and requiring that the parameter of a
message has this annotation when sending to a transferable object.

Another simplification in these formalisms is the absence of
fields in passive objects. If objects could refer to other
objects, a transferable object would not be allowed to refer to
actor-local passive objects, as these references would violate
actor isolation if the object was transferred. In Kappa, the type
system tracks if an object is (thread- or) actor-local, so
preventing transferable objects from referring to such objects is
straightforward.
Transferable objects may safely refer to
objects of active type (actors or transferable/bestowed objects)
or passive objects that are encapsulated by the transferable
object itself. Notably, with bestowed objects no such restriction
is necessary; as a bestowed object is only ever operated on by its
owner, it may only ever get passive references to objects with the
same owner.

\subsection{Private Message Queues}

\begin{figure}[t]
\begin{minipage}[t]{.45\textwidth}
\vspace{0pt}
\begin{tabular}{rcl}
$e$ & $::=$ &  $x$
           $|$ $e$ $e$
           $|$ $e\c{!}v$
           $|$ $e.$\c{mutate()}
           $|$ \c{new} $\tau$\\
          &$|$& \c{bestow} $e$
           $|$ \c{atomic} $e$
           $|$ \c{release} $e$
           $|$ $v$ \\
$v$ & $::=$ &  $\lambda x : \tau . e$
           $|$ $()$
           $|$ $\mathit{id}$
           $|$ $\iota$
           $|$ $\iota_{\mathit{id}}$ \\
\end{tabular}
\end{minipage}
\hfill
\begin{minipage}[t]{.45\textwidth}
\vspace{0pt}
\begin{tabular}{rcl}
$\tau$ & $::=$ & $\alpha$
           $|$ $\mathsf{p}$
           $|$ $\tau \rightarrow \tau$
           $|$ \texttt{Unit} \\
$\alpha$ & $::=$ & $\mathsf{c}$
           $|$ $\mathbf{B}(\mathsf{p})$ \\
\end{tabular}
\end{minipage}
\caption{\FigLabel{private-syntax} The syntax of a simple lambda calculus
  with actors with private message queues.}
\end{figure}

\begin{figure}[t]
\drules[e]
  {$\Gamma \vdash e : \tau$}
  {Expressions}
  {atomic
  ,release
  }
  \caption{\FigLabel{private-static} New static rules for dealing
    with private message queues.}
\end{figure}

\begin{figure}[t]
\drules[eval]
  {$\langle \mathcal{M}; H \rangle \overset{id}{\hookrightarrow} \langle \mathcal{M}'; H' \rangle$}
  {Evaluation}
  {actorXXprivXXrun
  ,actorXXprivXXend
  }
\caption{\FigLabel{private-dynamic1} New dynamic rules for dealing with private message queues (1/2).}
\end{figure}

\begin{figure}[p]
\drules[eval]
  {$\mathit{id} \vdash \langle \mathcal{M}, H, e \rangle \hookrightarrow \langle \mathcal{M}', H', e' \rangle$}
  {Evaluation of expressions}
  {atomicXXactor
  ,atomicXXbestowed
  ,atomicXXactorXXfail
  ,atomicXXbestowedXXfail
  ,releaseXXactor
  ,releaseXXbestowed
  ,releaseXXactorXXfail
  ,releaseXXbestowedXXfail
  ,sendXXactorXXatomic
  ,sendXXbestowedXXatomic
  }
  \caption{\FigLabel{private-dynamic2} New dynamic rules for
    dealing with private message queues (2/2). The left column
    handles rules for actor targets and the right column handles
    rules for bestowed targets. The helper function $\mathcal{C}$
    extracts the conversation map $C$ from an actor. The
    evaluation context $E$ is also extended with two cases:
    $E[\bullet] ::=~ \dots ~|~
    $\c{atomic}$~\bullet ~|~ $\c{release}$~\bullet$.}
\end{figure}

\begin{figure}[t]
  \centering
\drules[wf]
  {$H \vdash \mathcal{M} \qquad \vdash H \mathcal{M}; H; id \vdash (\iota, L, C; Q, e) \qquad \mathcal{M}; H; L; id \vdash Q$}
  {Well-formedness}
  {queueXXmap
  ,heapXXpriv
  ,actorXXpriv
  ,queueXXatomic
  ,queueXXend}

\caption{\FigLabel{private-wf} New and changed well-formedness
  rules for dealing with private message queues. }
\end{figure}

The calculus presented in \SecRef{formalism} encodes atomic
operations by batching several operations in a single message.
While this gives some control over the interleaving of messages,
it does not allow the sending actor to react to intermediate
results of the atomic operation. % It also does not handle atomic
% operations involving several actors.,
%
For example, an actor interacting with a list may want to
atomically iterate over the list, operating on each element along
the way, without allowing interleaved operations on the list in
the meantime.

This section extends our core calculus with support for
\emph{private message queues}. An actor can start a private
``conversation'' with another actor by creating a new queue and
passing it as a message. When an actor receives a new queue it
will read messages from this queue until it receives a message to
end the conversation. Messages sent by other actors will still be
enqueued in the actor's public queue.

On the surface level, two dual operations are added:
\c{atomic}$~e$ begins a new atomic operation, and \c{release}$~e$
finishes it (\cf{} \FigRef{private-syntax}). The target of
\c{atomic} or \c{release} can be actors or bestowed objects (\cf{}
\FigRef{private-static}). In the latter case, the owner of the
bestowed object gets sent the private queue. The two operations do
not have to appear in the same procedure body, and the language
does not enforce that an \c{atomic} expression has a matching
\c{release}. In a real programming language, using a blocked
construct is probably more practical---this is how \c{atomic} is implemented in Encore.

Dynamically, we need three extensions to the configuration. First,
each actor is extended with a map $C$ of its ongoing
conversations. $C$ maps actors $id$ to queue identifiers $q$.
Second, a global map $\mathcal{M}$ maps queue identifiers $q$ to
queues and the owner of the queue (the actor who is reading from
it). Third, a message can now be a request for starting a private
conversation using queue $q$ ($\mathbf{At}(q)$), a request for
ending a conversation ($\mathbf{End}$), or a normal message
modelled as anonymous function as before.

\FigRef{private-dynamic1} shows the two new rules for how actors
react to atomic messages. If the head of the actor's public queue
is $\mathbf{At}(q)$, the private queue corresponding to $q$ is
looked up in $\mathcal{M}$ and the first message from this queue
is run, leaving the public queue intact \RN{eval-actor-priv-run}.
If the head of the private queue is $\mathbf{End}$, the private
queue is dropped from $\mathcal{M}$ and the $\mathbf{At}(q)$
message is popped from the public queue, allowing the actor to
continue reading messages normally \RN{eval-actor-priv-end}.

\FigRef{private-dynamic2} shows the rest of the new dynamic rules.
Since we handle starting and finishing of atomic operations, on
both actor and bestowed objects, and these operations can succeed
or fail, there is a large number of new rules. The rules in the
left column all handle actor targets, and the rules in the left
column all handle bestowed targets.

Initiating an atomic operation with actor $id'$ creates a fresh
queue identifier $q$ and updates the current actor so that its
conversation map $C$ maps $id'$ to $q$. The global map is updated
so that $q$ maps to an empty queue belonging to $id'$, and an
atomic request $\mathbf{At}(q)$ is enqueued to the public queue of
$id'$ \RN{eval-atomic-actor}. If the current actor already has a
private conversation with $id'$---that is, $id'$ is already in
$C$---the expression is a no-op \RN{eval-atomic-actor-fail}. The
rules for a bestowed target is symmetric \RN{eval-atomic-bestowed-*}.

Finishing an atomic operation with actor $id'$ looks up the queue
identifier of the conversation with that actor in $C$ and appends
an $\mathbf{End}$ message to the end of the corresponding private
queue. Since this denotes the end of the conversation, the mapping
from $id'$ is dropped from $C$ \RN{eval-release-actor}. If there
is no ongoing conversation with $id'$---that is, $id'$ is not in
$C$---the expression is a no-op \RN{eval-release-actor-fail}.
Again, the rules for a bestowed target is symmetric
\RN{eval-release-bestowed-*}.

Finally, sending a message to an actor $id'$ with whom the current
actor has a private conversation looks up the queue identifier of
this conversation in $C$ (extracted from the actor using the helper function $\mathcal{C}$) and adds the message to the corresponding
private queue in $M$ \RN{eval-send-actor-atomic}. The case for a
bestowed target is symmetric \RN{eval-send-bestowed-atomic}. The
rules for normal message sends (not shown here) are extended with
a premise that there is no ongoing private conversation with the
receiving actor.

\FigRef{private-wf} shows the new and changed well-formedness
rules. The queue map $\mathcal{M}$ is well-formed if for each
mapping $q \mapsto (Q, id)$, the private queue $Q$ is well-formed
with respect to actor $id$ and does not contain any requests for
new private conversations (any such new requests would arrive in
the public queue of the actor). Additionally, if there is an
$\mathbf{End}$ in the queue, there can be no actor in the heap
that has a private conversation through the queue identified by
$q$, since this actor has just ended the conversation
\RN{wf-queue-map}.

The rule for well-formed heaps is extended with a premise stating
that the ranges of all actors' private conversations must be disjoint, meaning that no
two actors may send messages to the same private queue
\RN{wf-heap-priv}. The rule for well-formed actors states that
each entry in the conversation map has a corresponding entry in
the global queue map. The public queue of an actor must not
contain $\mathbf{end}$ messages (since these are necessarily sent
to the private queue), and all requests for atomic operations must
use distinct queue identifiers \RN{wf-actor-priv}. There are two
new well-formedness rules for queues. A request for an atomic
operation must have a corresponding private queue waiting in the
global queue map \RN{wf-queue-atomic}. $\mathbf{End}$ messages are
always well-formed \RN{wf-queue-end}.

To show that our private mailboxes are sound we define an informal
property:

\begin{quote}
  \textbf{Atomicity}: Once a private conversation between actors
  $A$ and $B$ has been initiated, actor $B$ will only read
  messages sent by actor $A$, up until and including the
  $\mathbf{End}$ message.
\end{quote}

The key rules for showing that atomic operations run without
interleaving are \ARN{wf-heap-priv}, which states that two
different actors cannot have a private conversation with the same
actor (preventing other actors from ``hijacking'' the private
queue that $A$ is writing to), and \ARN{wf-actor-priv}, which
states that an ongoing private conversation with actor $id'$ has a
corresponding queue belonging to $id'$ in the global queue map
(guaranteeing that messages sent from $A$ to $B$ will indeed end
up in the mailbox that $B$ is reading from). From the dynamic
rules \ARN{eval-actor-priv-*} it is easy to see that an actor will
keep reading from the same private message queue until it contains
an $\mathbf{End}$ message.

The formulation of the remaining meta-theoretic properties requires adapting
the statement of progress to include the state when an actor is
waiting for a message to arrive to its current private queue.
After this adaptation, the same soundness properties hold for this
variation of the formalism as well. The mechanised version of the
semantics are $\approx$670 lines of Coq, and the proofs are
$\approx$4000 lines with the same $\approx$600 lines of list
lemmas and tactics as for the other calculi \cite{bestow-repo}.
The increase in size compared to the other mechanisations is
explained by the large number of cases for evaluation as seen in
\FigRef{private-dynamic2}. There is room for reducing the code
size by applying more automation to similar cases of the proofs.

\subsubsection{Design Considerations for Private Queues}

The way private mailboxes are formalised here purposely have them
behave similar to locks in mainstream programming languages. The
\c{atomic} and \c{release} keywords work like acquire and release
on a lock, and all operations carried out by an actor between
calls to \c{atomic} and \c{release} enjoy absence of interleaving.

This is different from the implementation of \c{atomic} in Encore,
where a new handle to the target actor is created whose type
expresses that messages sent to this actor will not be interleaved
with messages from other actors. This handle is statically
restricted from being passed to other actors and cannot survive
its corresponding atomic block (it is effectively borrowed (\cf{}
\cite{Boyland,TheWrigstad})). This forces code to explicitly
propagate this handle to everywhere where the handle is needed.

Allowing such a handle to be shared with other actors is possible,
as is bounding the duration of the sharing to the static scope of
the atomic block. This would extend the number of sources of
messages in a private conversation, and re-introduce interleaving
in a controlled fashion. We have yet to find compelling enough
reasons to support this in real programs.

\section{Exploring Bestow and Atomic---Case Studies}
\SecLabel{case-studies}

In this section, we report on three case studies using bestow
and atomic to implement distributed hash tables, distributed
graphs, and atomic money transfer. These case studies highlight
the usefulness of bestow and atomic and some of their
properties, and also serve as more compelling motivating examples
(although less pedagogic) than the list example of
\SecRef{example}.

\subsection{Case Study: Distributed Hash Tables}

The Encore programming language was recently extended with support
for ``locally distributed'' arrays and hash tables. These data
structures are backed by several actors to enable true parallel
operations on disjoint parts. Distributing a hash table $T$ over
$N$ different actors allows up to $N$ operations on $T$ to be
carried out in parallel, provided that there are at least $N$
cores where the actors are scheduled in parallel\footnote{This is
  ultimately a decision made by the run-time system, and depends
  on the available cores, system load, etc.}.

Encore's distributed hash tables are implemented purely as a
library construct. Each client of a hash table holds a single passive object which acts as a proxy for the
hash table and to this end keeps a map from hash ranges to actors
that it uses to delegate operations to the correct actor.
Distributing this ranges-to-actors map across all proxies is key
to the parallel performance.

Given the aforementioned map, puts and gets are straightforward to
implement. A more complex operation is \emph{rehashing}, \ie{}
updating the map, possibly involving changing the number of actors
backing the data structure in the process. Rehashing involves
stopping all operations in all actors backing the data structure,
reconfiguring the map, and resume operations with the new map.
Since clients may continue to send messages using an outdated map
during rehashing, requests must be buffered until the new map
is in place, and possibly rerouted to another actor. The top right
of \FigRef{rehash} shows excerpts of a \c{put} method that
buffers messages during rehashing.

\FigRef{rehash} (left) shows a high-level implementation of the
rehashing function. The important lines with respect to bestow are
16 and 17, iterating over all hash-table proxies and updating
their ranges-to-actors maps. The type of each \c{p} is
\c{B(HashMap)} meaning each call to update is wrapped in a closure
which is sent to the containing actor. This design requires that
all proxies are known by the distributed hash table, which is
easily handled \eg{} in a constructor and/or factory method. For
simplicity, we omit a versioning scheme checking that the map used
by a proxy is up to date (for example, checking that the
identity of the client's map was the same as the actor's current
map). This handles the case when calls are made on an outdated
proxy after rehashing is finished.

Without the ability to create bestowed references, we are
left with two---problematic---possibilities to implement rehashing:

\begin{description}
\item[Polling for changes to the ranges-to-actors map]

  In this scenario, proxies would continuously poll for changes to
  the map. Since polls cross actor boundaries, they would be
  asynchronous increasing the latency of hash table operations,
  possibly causing client actors to block waiting for polling
  responses.

\item[Forcing actors using hash-tables to implement support methods]

  In this scenario, actors using hash-tables would be required to
  implement a special support method for updating client proxies.
  This has the downsides of pushing internal implementation detail
  out into the interface, not just for the actors themselves, but
  any library used inside the actor must somehow be interfaced with
  by the top-level to support updating ranges-to-actors maps.

\end{description}

As \FigRef{rehash} shows, it is possible to avoid these problems
through bestowed references as these allow us to \emph{push}
changes directly to the hash table proxies---without any special
requirements on the clients holding them.

\begin{figure}[t]
  \begin{minipage}[t]{.48\linewidth}
\begin{codep}
def rehash() : unit
  -- Send a stop message to all actors
  val result = for a <- this.actors do
                 a ! stop(new_map)
  -- Create a new actor map
  val new_map = this.create_new_map()
  -- Block until all actors have stopped
  for r <- result do
    r.get()
  -- Update the list of actors
  this.actors = new_map.values()
  -- Start actors up using new map
  for a <- this.actors do
    a ! start(new_map)
  -- Update all proxies' map
  for p <- this.proxies do
    p ! update(new_map)
\end{codep}
  \end{minipage}
  \begin{minipage}[t]{.48\linewidth}
\begin{codep}
def put(key : k, value : v) : unit
  if this.is_rehashing then
    this.buffer.append(
      (this.PUT_OP, key, value))
  else
    ... -- omitted

def start(map : k -> HashActor[k, v]) : unit
  this.is_rehashing = false
  for op <- this.buffer do
    match op with
      (this.PUT_OP, key, value) =>
        map(key) ! put(key, value)
      ...
\end{codep}
  \end{minipage}
  \caption{Using bestow to implement rehashing (simplification).}
  \label{fig:rehash}
\end{figure}

An obvious extension to the atomic construct is the ability to
atomically operate on a number of subjects (actors in the current
example). In the case of a statically known number of subjects, it
is easy to implement \c{atomic (a, b) ...} as syntactic sugar
for nested atomic blocks. In the case of a dynamic number,
something similar can be done manually through recursion. If we
imagine a powerful atomic block able to suspend a number of
actors before performing its operations on all of them, and
then releasing them, the \c{rehash()} function of
\FigRef{rehash} could be greatly simplified:

\begin{codep}
def rehash() unit
  val new_map = this.create_new_map()
  atomic this.actors do -- operate atomically on all actors in this set
    for a <- this.actors do
      a ! update(new_map)
  for p <- this.proxies do
    p ! update(new_map)
\end{codep}

Notably this implicitly takes care of any buffering of messages
sent to actors backing the hash table while rehashing is taking
place. Furthermore, the aforementioned versioning scheme dealing
with messages sent via proxies whose maps are out of date could be
eliminated if lines 6 and 7 could be made part of the atomic block,
but would only be advisable if the number of proxies was very
small.

\subsection{Case Study: Distributed Graphs}

Just like distributing a hash table over a collection of actors,
there are many cases where it is appropriate to distribute a graph
over a collection of actors. A simple representation of a graph is
an object structure where objects representing nodes model edges
using references, keeping a separate list of weights for edges.
When distributing such a model over several actors, different
actors own different nodes in the graph, and thus edges must
sometimes connect nodes across different actors. Here, bestowed
references can be used to abstract whether an edge is inter-actor
(meaning it points to a node inside the same actor) or
intra-actor.

\begin{figure}[t]
  \centering
  \includegraphics[width=.9\textwidth]{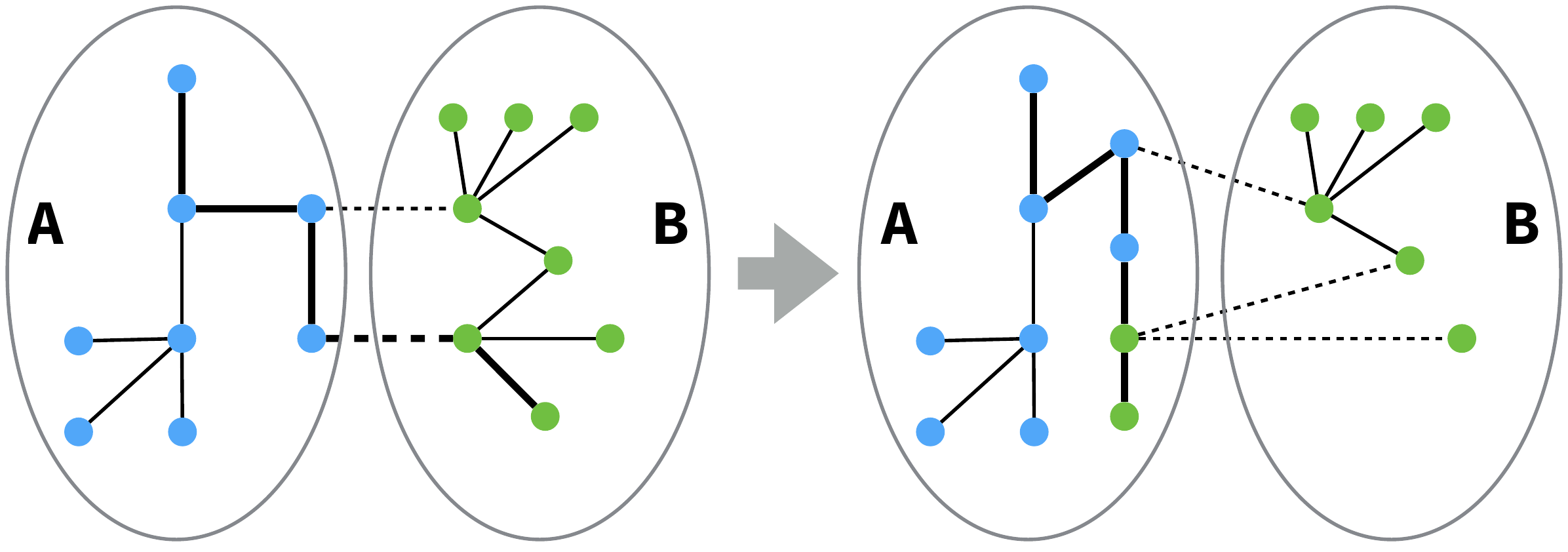}
  \caption{Load-balancing of passive nodes in a distributed graph.
    The black circles A and B are actors backing a graph
    consisting of blue and green nodes (to highlight original
    ownership), implemented as passive objects. Lines denote
    edges, implemented as references. Dashed lines highlight
    inter-actor references.}
  \label{fig:load-balancing}
\end{figure}

A straightforward implementation of Dijkstra's shortest path
algorithm can be applied with bestowed edges, but will suffer from
added latency due to many lookups of edges or weights changing to
asynchronous message sends. This highlights the importance of
syntactically highlighting asynchronous sends with \c{!}, and
requiring the act of bestowing to be explicit. Consider for
example the following code, which (possibly) updates the shortest
paths to the nodes reachable from the current node:

\begin{codep}
for (n, dist) <- src ! edges() do
  unless visited.contains(n) then
    val new_distance = distances(src) + dist
    if new_distance < distances(n) then
      distances(n) = new_distance
      predecessors(n) = src
\end{codep}

On Line 1, we obtain the edges and weights for the neighbours of
node \c{src}. If \c{src} points to another actor $o$, this lookup
will involve an asynchronous call to $o$ to obtain the result. If
\c{src} is local to the current actor, this operation can be
converted to a fast synchronous lookup. In this code, we only use
the identity of \c{n}, which is a fast operation regardless
whether \c{n} is a bestowed reference or not.

\FigRef{load-balancing} depicts a graph distributed over two
actors. The thick path denotes a ``hot path'' in the system,
meaning a path that is traversed often. In a system that is able to
load-balance passive objects as discussed in
\SecRef{load-balancing} and formalised in \SecRef{formal2},
traversing the hot path could lead to moving the green objects on
the hot path from actor $B$ to actor $A$, allowing $A$ to operate
synchronously on all the objects in the path, without any latency added
by crossing a dashed edge. The figure also illustrates how
load-balancing might lead to increasing inter-actor references.
The performance impact of increased number of inter-actor
references naturally depends on how frequently they are
dereferenced.

\subsection{Case Study: Money Transfer}
\label{sec:money-transfer}

Money transfer across bank accounts is a classic example of
atomicity. Imagine actors acting as banks, and each bank holding
passive objects as bank accounts. Foregoing whether unrestricted
external access to a bank account is advisable from a security
perspective, bestowed references allows exposing the interfaces of
individual bank accounts outside of the banks, and also the
construction of an atomic transfer operation even if the banks
themselves do not support it. The following function uses a
combination of bestowed pointers and atomic blocks to implement
transfer in a straightforward fashion:

\begin{codep}
def transfer(amount : int, from : B(Account), to : B(Account)) : unit
  atomic from do
    atomic to do
      if from ! withdraw(amount) then
        to ! deposit(amount)
\end{codep}

Notably, providing syntactic sugar for the nested atomic
blocks is straightforward. When this block is executed, if the
accounts are owned by the same actor, the
whole content of the atomic block can be sent to the owning actor
to be performed there in a synchronous fashion. Outside of this
special case, because we are nesting atomic blocks, external messages to
the actors referenced by \c{from} and \c{to} will not be received
until \emph{after} Line 4. Thus, no other actor will be able to
witness that \c{amount} money is ``missing'' in the system,
between Lines 3 and 4.

Notably, a single atomic block \emph{can} be strong enough to
support a limited form of inter-actor atomicity for two actors, if
the current actor is one of them, with a small modicum of extra
work for the programmer:

\begin{codep}
def transfer(amount : int, from : Account, to : B(Account)) : unit
  atomic to do
    if from.withdraw(amount) then
      val done = to ! deposit(amount)
      done.get()
\end{codep}

On Line 3, \c{withdraw()} can be called synchronously because
\c{from}'s type captures that it is a local object. Thus, external
actors' inability to witness the change before the end of the
\c{transfer()} method arises naturally from the
single-thread-of-control invariant of actors. On Line 4, we
capture the future return value of \c{deposit()} and block on its
return on Line 5. Thus, the transfer method will not finish until
the money has been deposited in the target account.

Note that because the type of \c{from} is not bestowed, this
version of \c{transfer()} cannot be called externally by other
actors. For that to be possible, there needs to be some way for
external actors to identify a particular account, such as a unique
account number.

\section{Implementation in Encore}
\SecLabel{implementation}

A prototypical implementation of bestow and atomic exists
in Encore, and is currently being hardened to become part of the
main branch. The prototype has been instrumental in understanding
the performance implications of the two constructs, especially in
relation to Encore's support for fully concurrent garbage
collection \cite{Art:ORCAOOPSLA}. A more comprehensive account of the
implementation aspects can be found in \cite{JoelBSC}.

We extend each actor class with an implicit method \c{perform}
which takes a function, applies it to the \c{this} of the
receiver, and returns the result wrapped in a future. A bestowed
reference is logically implemented as a pair of values,
\c{owner} and \c{object}. A message send \c{x ! m()} on a
bestowed reference is translated into the message send \c{x.owner !
  perform((}$\lambda$ \c{\_ . x.object.m()))}.

The atomic block is implemented using the private mailbox
semantics sketched in \SecRef{atomic} and formalised in \SecRef{formal2}. An
atomic block is turned into a pair of matching \c{acquire} and
\c{release} operations, where the former creates and installs the
private mailbox and returns an atomic reference, and the latter
sends the termination message that reinstates the mailbox at the
start of the atomic block. Similar to a bestowed reference, an
atomic reference is logically implemented as a pair of values
\c{target} and \c{mailbox}, and a message send \c{x ! m()} on an
atomic reference uses a special message sending primitive that
directs the message to \c{mailbox} instead of the normal mailbox.

\begin{minipage}{.2\textwidth}
\begin{code}
atomic x do
  x ! foo(42)
  x ! bar(-42)
\end{code}
\end{minipage}
$\implies$
\begin{minipage}{.7\textwidth}
\begin{code}
let x' = acquire x in
  x' ! foo(42)
  x' ! bar(-42)
  release x'
\end{code}
\end{minipage}

% \begin{code}
% atomic from
%   atomic to
%     val success = from ! withdraw(amount)
%     if success then
%       to ! deposit(amount)
% \end{code}

% \noindent
% In this example, we cannot batch the messages as they are sent to
% different receivers. As an alternative, we are envisioning an
% implementation that allows temporarily switching the message queue
% of an active object for one that is only accessible to the caller.
% This prevents the active object from processing messages from
% other senders for the duration of the \c{atomic} block. The
% \c{atomic} block above could then (logically) be translated into
% the following code:

% \begin{code}
% // Exchange message queues
% val oldFromQueue = from ! getQueue()
% val newFromQueue = new MessageQueue()
% from!setQueue(newFromQueue)
% val oldToQueue = to ! getQueue()
% val newToQueue = new MessageQueue()
% to!setQueue(newToQueue)

% // The actual computation
% val success = newFromQueue.enqueue(withdraw, amount)
% if success then
%   newToQueue.enqueue(deposit, amount)

% // Restore the old queues when done
% from ! setQueue(oldFromQueue)
% to ! setQueue(oldToQueue)
% \end{code}

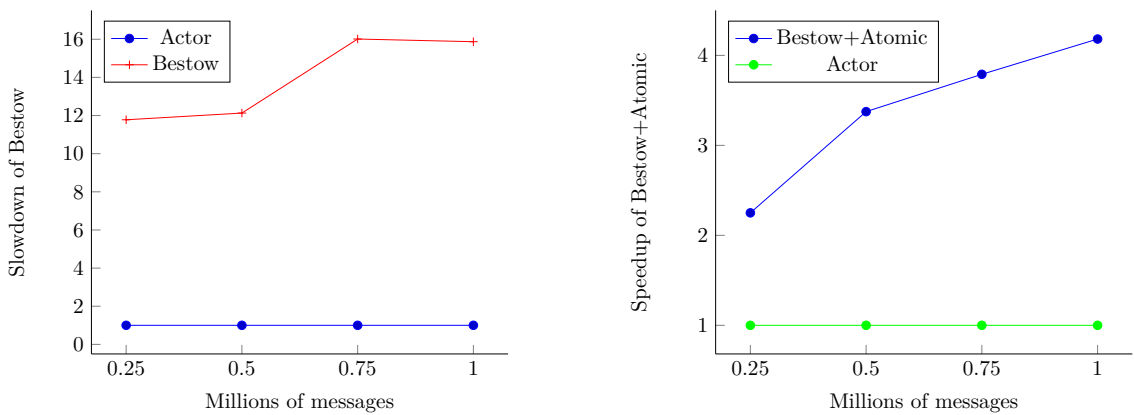
\begin{figure}[ht!]
\hspace*{-1cm}
%\begin{widepage}
\centering
\begin{tikzpicture}[scale=0.8]
\begin{axis}[
  xlabel=Millions of messages,
  ylabel=Slowdown of Bestow,
  xtick = {0, 0.25, 0.5, 0.75, 1},
  ytick = {0, 2, 4, 6, 8, 10, 12, 14, 16, 18},
  legend pos=north west,
  axis x line*=bottom,
  axis y line*=left]
\addplot table [y=Actor, x=Seconds]{bestowOverhead2.dat};
\addlegendentry{Actor}
\addplot+[red,mark=+,mark options={fill=red}] table [y=Bestow, x=Seconds]{bestowOverhead2.dat};
\addlegendentry{Bestow}
\end{axis}
\end{tikzpicture}
\qquad\qquad
\begin{tikzpicture}[scale=0.8]
\begin{axis}[
  xlabel=Millions of messages,
  ylabel=Speedup of Bestow+Atomic,
  legend pos=north west,
  xtick = {0, 0.25, 0.5, 0.75, 1},
  ytick = {0, 1, 2, 3, 4, 5, 6},
  axis x line*=bottom,
  axis y line*=left]
\addplot table [y=Actor, x=Seconds]{atomicOverhead.dat};
\addlegendentry{Bestow+Atomic}
\addplot+[green,mark=*,mark options={fill=green}] table [y=AtomicBestow, x=Seconds]{atomicOverhead.dat};
\addlegendentry{Actor}
\end{axis}
\end{tikzpicture}
\caption[Comparison: Ping benchmark with bestow and atomic]{Overhead of message sends on bestowed objects, and using atomic closure to reduce the impact of the overhead.}
\label{fig:ping-overhead}
%\end{widepage}
\end{figure}

In previous work \cite{Art:Bestow, JoelBSC}, we explored an
implementation of atomic with coalescing semantics. While this is
less general than the private mailbox semantics, our original
assumption was that it might serve as an optimisation in simple
uses of atomic. However, an implementation of bestow close to the
formal semantics of coalescing (\cf{} \SecRef{formalism}) involves
creation of closures passed from the operating actor to the owner
of the passive object. Without special optimisation, the overhead
of closure creation, and garbage collection of the closure value
is significant. We illustrate this in \FigRef{ping-overhead}
(left) through a modified version of the Ping benchmark
from the Computer Language Benchmark Game \cite{bench_game} where two actors are
involved in a ping--pong exchange of messages. The figure shows a
fairly constant slowdown for the bestowed version of
11--16$\times$ over the non-bestowed counterpart. The main source
of this overhead is fully concurrent garbage collection---the
closure is being created on the local heap of one actor and then
sent to another, which must eventually inform the creating actor
when the closure is no longer needed. Passing closures by copy is
likely to reduce this overhead. This benchmark clearly
demonstrates that programs mostly operating on bestowed pointers
might see a noticeable slowdown.

In contrast, the right-hand-side of \FigRef{ping-overhead} shows
that the combination of bestow and atomic can outperform
the actor-only solution. By wrapping the loop containing the
message sends in an atomic, the entire loop can be transferred
to the receiver where it can be run synchronously. This reduces
the number of asynchronous message sends, which involve expensive
\c{CAS} operations on mailboxes, and enables compile-time
optimisations such as inlining.

In \SecRef{grouping-semantics}, we noted that atomic blocks
add structure to actor-programs. If desired, this can be leveraged
in scheduling. When executing \c{atomic a do ...}, if the actor
denoted by \c{a} is \emph{idle}, meaning its mailbox is empty, it
is possible to turn the asynchronous operations inside \c{...}
synchronous, \ie{} turning \eg{} \c{a ! ping()} into \c{a.ping()}.
This avoids context switching, enables inlining as well as other
optimisations such as removing future indirection which are costly
in Encore. Static analysis can be used to determine whether
removing asynchrony \emph{may} also remove possible parallel
gains. Making asynchronous operations synchronous allows mixing
the methods of several actors on the same stack frame because
control flow is guaranteed to match the order that frames are
pushed onto and popped from the stack.

\section{Related Work}
\SecLabel{related}

An important property of many actor-based systems is that a single
actor can be reasoned about sequentially; messages are exchanged
concurrently but executed sequentially by the receiving actor. For
this property to hold, actors often rely on \emph{actor isolation}
\cite{kilim}, \ie{} that the state of one actor cannot be accessed
by another. If this was the not the case, concurrent updates to
shared state could lead to data-races, breaking sequential
reasoning.

Existing techniques for achieving actor isolation are often based
on restricting aliasing, for example copying all data passed
between actors \cite{erlang}, or relying on linear types to
transfer ownership of data \cite{encoreSFM, ponyAgere, kilim,
  haller2010}. Bestowed objects offer an alternative technique
which relaxes actor isolation and allows sharing of data without
sacrificing sequential reasoning. Combining bestowed objects with
linear types is straightforward and allows for both ownership
transfer and bestowed sharing between actors in the same system.

Miller \etal{} propose a programming model based on function
passing, where rather than passing data between concurrent actors,
functions are sent to collections of stationary and immutable data
called \emph{silos} \cite{miller}. Bestowed objects are related in
the sense that sharing them doesn't actually move data between
actors. In the function passing model, they could be used to
provide an interface to some internal part of a silo, but
implicitly relay all functions passed to it to its owning silo.
While the formalism in \SecRef{formalism} also works by passing
functions around, this is to abstract away from unimportant
details, and not a proposed programming model.

References to bestowed objects are close in spirit to remote
references in distributed programming or eventual references in E
\cite{RobustComposition}. In the latter case, the unit of
encapsulation, \eg{} an actor or an aggregate object protected by
a lock, acts similar to a Vat in E, but with an identifiable
boundary and an identity with an associated interface.
By bestowing and exposing sub-objects, a unit of encapsulation can
safely delegate parts of its interface to its inner objects, which
in turn need not be internally aware of the kind of concurrency
control offered by their bestower.

The Emerald language \cite{black2007development} has an entire
``locatics'' system for expressing object placement and mobility
combined with an object-based language. Very early, Emerald
introduced \emph{call-by-move} which allowed parameters of a call
to a remote object to be moved to the target's location
\cite{jul1988fine}.

Any actor system which supports sharing of mutable state could
implement individual bestowed objects by using a wrapper with the
same interface as the bestowed object, but where all operations
are delegated to some actor. Similarly, in a language with locks,
a wrapper object could grab a specified lock before calling the
underlying methods.
In both cases, this comes at the cost of explicitly creating
wrapper objects for all bestowed objects, and requires that the
programmer never directly accesses the object inside the wrapper.
For actor systems, bestowed references make the most sense when
messages are handled in sequence. In languages where messages are
explicitly and selectively received by each actor, the owning
actor would have to regularly check for incoming delegated
operations.

X10 introduces a language construct that is called atomic for
executing multiple operations in sequence without interleaving
\cite{Art:X10}, but with different semantics and notably no
support for side effects or blocking operations.

Ibrahim \etal{} introduce the concept of Remote Batch Invocation
for combining operations to be run remotely in a distributed
system (in Java)~\cite{batching}. This is similar to the coalescing
semantics of atomic blocks, but more advanced since parts of a batch
may consist of local computations, allowing the client to react to
the results of the remote operations inside the batch, while still
only making a single round-trip to the server. The private mailbox
semantics of atomic blocks is more flexible than Remote Batch
Invocation, but would be more expensive in a distributed setting
where each remote call is costly.

The coalescing semantics of atomic blocks can be implemented in
any actor language with closures by sending a closure performing
the desired operations in a message. This requires that the
receiving actor has support for running arbitrary closures, which
complicates reasoning about the behaviour of that actor (in our
Encore implementation, the \c{perform} method is only ever called
implicitly as an effect of an atomic block).
Languages with selectively receives can encode the kind of
``private conversation'' between two actors, offered by the
private mailbox semantics of atomic blocks, but requires that the
receiving actor is implemented with support for atomically
processing exactly the messages that the client wanted to send.

To the best of our knowledge, there are no other systems which
implicitly delegates locking based on ownership, as when bestowing
the internal state of a locked object.
Ownership types have been used to enforce strong encapsulation and
to allow a single lock to safely govern access to an entire object
aggregate, \eg{} in Universes for Race Safety~\cite{universes} or
Parameterized Race Free Java (PRFJ)~\cite{boyapati2002ownership}.
PRFJ additionally allows the programmer to specify a lock-order
and enforces that locks are taken in this order, preventing
deadlocks. Kappa does not attempt to prevent deadlocks, and
bestowed references to the private state of locked objects may be
subject to deadlocks.
Universes for Race Safety allow references with unknown owners,
possibly crossing encapsulation boundaries, but require explicitly
locking these objects before they are accessed. In contrast,
accessing a bestowed object only requires locking the \emph{owner}
object (which is done implicitly).

When using locks, the atomic block behaves like Java's
\c{synchronized} blocks, but Kappa additionally precludes
accidentally bypassing the lock of a locked object.
Locked objects in Kappa are protected by a readers-writer lock,
which allows a single writer, but multiple concurrent readers
(statically enforcing that readers will not cause mutation). A
bestowed reference into a locked aggregate which only exposes
reading operations can therefore acquire a reading lock, allowing
multiple concurrent readers through bestowed references to the
same aggregate.

\section{Discussion and Conclusion}
\SecLabel{discussion}

Although our formal description and all our examples focus on
actors, bestow also works with threads and locks. An object
protected by a lock can share one of its internal objects while
requiring that any interaction with this object also goes via this
lock.
We believe there is also a straightforward extension to software
transactional memory. In the future, we would like to study
combinations of these.

Bestowed objects let an actor expose internal details about its
implementation. Breaking encapsulation should always be done with
care as leaking abstractions leads to increased coupling between
modules and can lead to clients observing internal data in an
inconsistent state. The latter is not a problem for bestowed
objects however; interactions with bestowed objects will be
synchronised in the owning actor's message queue, so as long as
data is always consistent \emph{between} messages, we can never
access data in an inconsistent state (if your data is inconsistent
between messages, you have a problem with or without bestowed
objects).

Sharing bestowed objects may increase contention on the owner's
message queue as messages to a bestowed object are sent to its
owner. Similarly, since a bestowed object is protected by the same
lock as its owner, sharing bestowed objects may lead to this lock
being polled more often.
As always when using locks there is a risk of introducing
deadlocks, but we do not believe that bestowed objects exacerbate
this problem. Deadlocks caused by passing a bestowed object back
to its owner can be easily avoided by using reentrant locks (as
accessing them both would require taking the same lock twice).

When using locks, atomic blocks are very similar to Java's
\c{synchronized}-blocks. With actors, an atomic block groups
messages into a single message. For fairness, it may make sense to
only allow atomic blocks that send a limited number of
messages.

It is possible to synchronise on several locked objects by simply
grabbing several locks. Synchronising on several actors is more
involved, as it requires actors to wait for each other and
communicate their progress so that no actor starts or finishes
before the others. The canonical example of this is atomically
withdrawing and depositing the same amount from the accounts of
two different actors. Interestingly, if the accounts are bestowed
objects from the same actor (\eg{} some bank actor), this atomic
transaction can be implemented with the message batching approach
suggested in this paper. We leave this for future work.

Actor isolation is important to maintain sequential reasoning
about actors' behavior. By bestowing activity on its internal
objects, an actor can share its representation without losing
sequential reasoning and without bloating its own interface. With
atomic blocks, a client can create new behavior by composing
smaller operations. While bestowed references may not be very
efficient, due to their asynchronous nature, they can be used in
concert with atomic blocks to move entire operations closer to
data operated on, which can not only avoid the negative
performance implications of bestowed references, but lead to
performance improvements over pure actor-based solutions.

An earlier version \cite{Art:Bestow} of this work quipped ``actors
without borders'', in the sense of relaxing isolation-based
data-race freedom using our bestowed references. The same is true
for lock-based synchronisation---bestowed references allows
several objects to sit at the boundary of synchronisation, and
share a common lock. In some ways, this is similar to multiple
ownership \cite{MOJO,ombudsmen}, but extended to a concurrent
setting. By tracking bestowed-ness through types, operations on
resources owned by others are clearly visible at the use-site, but
need no forethought at the declaration-site: the bestowed objects
themselves do not need to know why access to them is safe, nor
even be aware of their sharing. They can simply trust the safety
of living in a world where borders are a means to an end, not an
end.

\end{document}

%% file: semantics.ott.tex
\newenvironment{ottdefnblock}[3][]{ \framebox{\mbox{#2}} \quad #3 \\[0pt]}{}

\newcommand{\ottafterlastrule}{\\}
\usepackage{latexsym}
\usepackage{amssymb}

\renewcommand{\bfdefault}{b}
\renewcommand{\mathbf}{\textbf}
\providecommand{\ottlinebreakhack}{}